\documentclass{aa}
\usepackage{graphicx}
\usepackage{txfonts}
\usepackage{natbib}
\bibpunct{(}{)}{;}{a}{}{,}

\begin{document}

\title{Radio perspectives on the Monoceros SNR G205.5+0.5}

\author{L. Xiao\inst{1}
        \and M. Zhu\inst{1}
}

\offprints{L. Xiao}

\institute{National Astronomical Observatories, Chinese Academy of
  Sciences, Jia-20, Datun Road, Chaoyang District, Beijing 100012,
  China\\ \email{xl@nao.cas.cn,mz@nao.cas.cn}
}

\date{Received / Accepted}

\abstract
{The Monoceros Loop (SNR G205.5+0.5) is a large shell-type supernova remnant located
in the Rosette Complex region. It was suggested to be interacting with the Rosette Nebula.
}
{We aim to re-examine the radio spectral index and its spatial variation over
the Monoceros SNR, and study its properties of evolution within the complex interstellar medium.}
{We extracted radio continuum data for the Monoceros complex region from the Effelsberg
21~cm and 11~cm surveys and the Urumqi 6~cm polarization survey. We used the new Arecibo
GALFA-HI survey data with much higher resolution and sensitivity than that previously
available to identify the HI shell related with the SNR. Multi-wavelengths data are
included to investigate the properties of the SNR.}
{The spectral index $\alpha$ ($S_{\nu}\propto\nu^{\alpha}$) averaged over
the SNR is $-0.41 \pm$0.16. The TT-plots and the distribution of $\alpha$
over the SNR show spatial variations which steepen towards the inner western
filamentary shell. Polarized emission is prominent on the western filamentary
shell region. The RM there is estimated to be about 30$\pm$77n~rad~m$^{-2}$, where the
n=1 solution is preferred, and the magnetic field has a strength of about 9.5~$\mu$G.
From the HI channel maps, further evidence is provided for an interaction
between the Monoceros SNR and the Rosette Nebula. 
We identify partial neutral hydrogen shell structures in the northwest region
at LSR velocities of +15~km s$^{-1}$ circumscribing the continuum emission.
The HI shell has swept up a mass of about 4000~M$_{\odot}$ for a distance of
1.6~kpc. The western HI shell, well associated with the dust emission, is found
to lie outside of the radio shell. We suggest that the Monoceros SNR is evolving within 
a cavity blown-out by the progenitor, and has triggered part of the star formation in
the Rosette Nebula.}
{The Monoceros SNR is interacting with the ambient interstellar medium with
ultra-high energy emission detected. Its interaction with the Rosette Nebula is further
supported by new evidence from HI data, which will help the investigation of the emission
mechanism of the high energy emission.}


\maketitle

\section{Introduction}
The Monoceros Nebula (SNR G205.5+0.5) is an extended filamentary object lying in the anti-center
complex region. The bright Rosette Nebula (SH 2-275) and HII region SH 2-273 are adjacent to the
edge of the southern and northeastern shell of the SNR. It was first recognized as a possible
supernova remnant by~\citet{d63} from radio observations at 237~MHz. More detailed observations
presented by~\citet{h68} confirmed this property by a non-thermal radio spectrum.

The SNR has been studied at low and intermediate radio frequencies by~\citet{h68, mh69}
and~\citet{dd75}. Observations at frequencies near 2700~MHz were made by~\citet{dcc72, md74, vk74}
and~\citet{ghs82}.~\citet{ghs82} found an averag spectral index $\alpha=-0.47\pm0.06$
from 111$-$2700~MHz radio data. They also presented the spatial spectral index distribution
at 178$-$2700~MHz and 611$-$2700~MHz, which was found to be steeper in the inner region. 
They also identified an HII region (G206.35+1.35) in the eastern periphery from its thermal spectrum.~\citet{ghr11} showed a new polarization observation at 4800~MHz. Combined this with
Effelsberg 21~cm and 11~cm data, they obtained a spectral index of $\alpha=-0.43\pm0.12$ for the
western shell. Early 11\ cm polarization observations were made by~\citet{md74}, but no significant polarized emission was detected. The 6\ cm polarization map showed that the magnetic
field is closely aligned with the filamentary shell.

The distance to the Monoceros SNR has been estimated to range from around 800~pc to about
1.6~kpc. ~\citet{o86} summarized all the available evidence and argued that the Monoceros SNR
is at the same distance as the Monoceros OB2 association of about 1.6~kpc. This is further
supported by the decimeter observations of the absorption of non-thermal emission from the
supernova remnant by the Rosette Nebula~\citep{o86}. The diameter of the SNR is then about 106~pc
(3.8$\degr$), and the age is about $1.5\times10^{5}$~yr.  A study of Einstein IPC data
~\citep{lns85,lns86} shows diffuse X-ray emission from the Monoceros SNR shock front, corresponding to the optical filaments. By fitting the data, they concluded that the SNR was
expanding in a low-density (0.003~cm$^{-3}$) medium, and has not yet entered the radiative
phase~\citep{lns86}. ~\citet{r72} identified a neutral hydrogen shell located outside of the
Monoceros optical filament with low resolution HI data.

The Monoceros SNR is suggested to be interacting with the Rosette Nebulae. In the SNR ridges
overlapping Rosette region,~\citet{fgo79} found broadened line widths of H109$\alpha$ and
much enhanced H$\alpha$ emission than average in the Rosette Nebula.  EGRET detected extended
$\gamma$-ray emission (3EG J0634+0521) from the region associated with the Monoceros/Rosette
Nebula in the energy range from 100~MeV up to 10~GeV~\citep{jbd97}. This emission was
interpreted as $\gamma$-rays from the decay of $\pi^{0}$'s produced by the interaction
of shock accelerated protons with the ambient matter. HESS found a source HESS J0632+057
(100~GeV $-$ 10~TeV) located at the rim of the western SNR shell~\citep{fhg08}, which seems
to be associated with 3EG J0634+0521 and an associated molecular cloud.
An observation carried out with the BeppoSAX satellite~\citep{kph99} discovered a hard spectrum
X-ray point source (SAX J0635+0533) within the 95\% probability circle of the EGRET detection,
which was later identified as a binary pulsar~\citep{cmn00}. ~\citet{trd03} postulated
that both the binary pulsar SAX J0635+0533 and the emission from the interaction region between
the expanding shell and the Rosette Nebula contribute to the EGRET $\gamma$-ray emission.

Although there is already a considerable body of radio observations on the Monoceros SNR,
we try to better delineate the extent of the SNR and the spatial spectral index variations using
new radio data with higher resolution. Also, a new high sensitivity, high resolution HI
data cube enables us to identify HI features morphologically associated with the SNR and
to discuss various evolutionary scenarios. In Sect.2, we describe the 11, 21, and 6~cm
radio continuum observations, as well as the HI data of the Monoceros SNR. The determination of
the flux densities and the spectral index distribution are presented in Sect.3. The polarization
properties and the magnetic fields in the northeast filamentary shell are estimated in Sect.4. 
The associated HI shell analysis is presented in Sect.5. Finally, we give a discussion and
summary in Sects.6 and 7, respectively.

\section{Data }
\subsection{Radio data}
The 21~cm and 11~cm radio continuum data presented here are taken from the Effelsberg Galactic
plane survey~\citep{rrf90,frr90}. The angular resolution of these surveys are 9.4$\arcmin$ and
4.4$\arcmin$, respectively. The 6~cm continuum and polarization data are extracted from the
Sino-German polarization survey of the Galactic plane~\citep{grh10} made with the Urumqi 25-m
telescope. The angular resolution is 9.5$\arcmin$ at 4.8~GHz. Detailed system and observation
parameters and the data reduction procedure are described in these related papers.

\begin{figure}[!hbt]
\begin{center}
\includegraphics[angle=-90,width=0.48\textwidth]{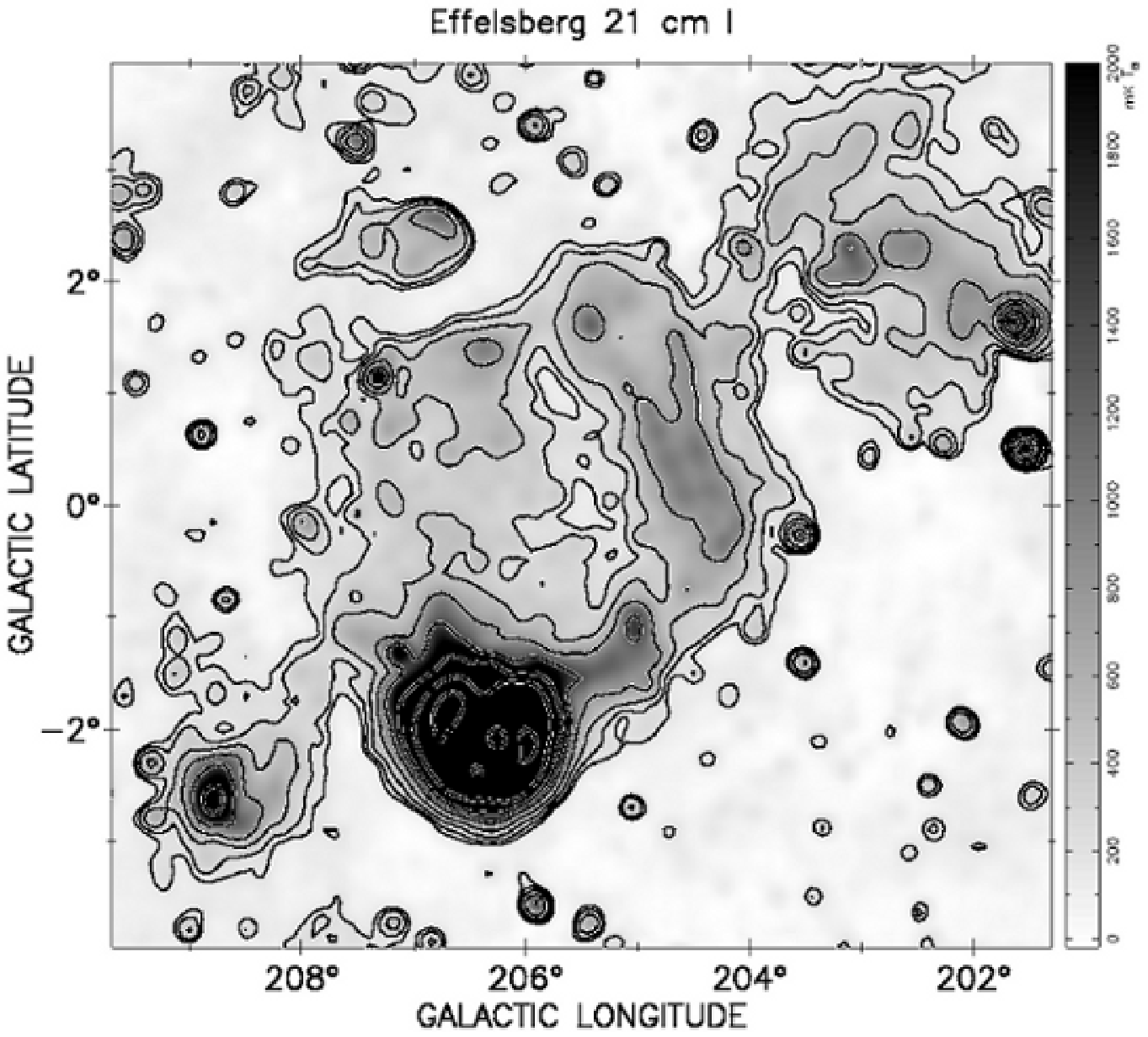}
\includegraphics[angle=-90,width=0.48\textwidth]{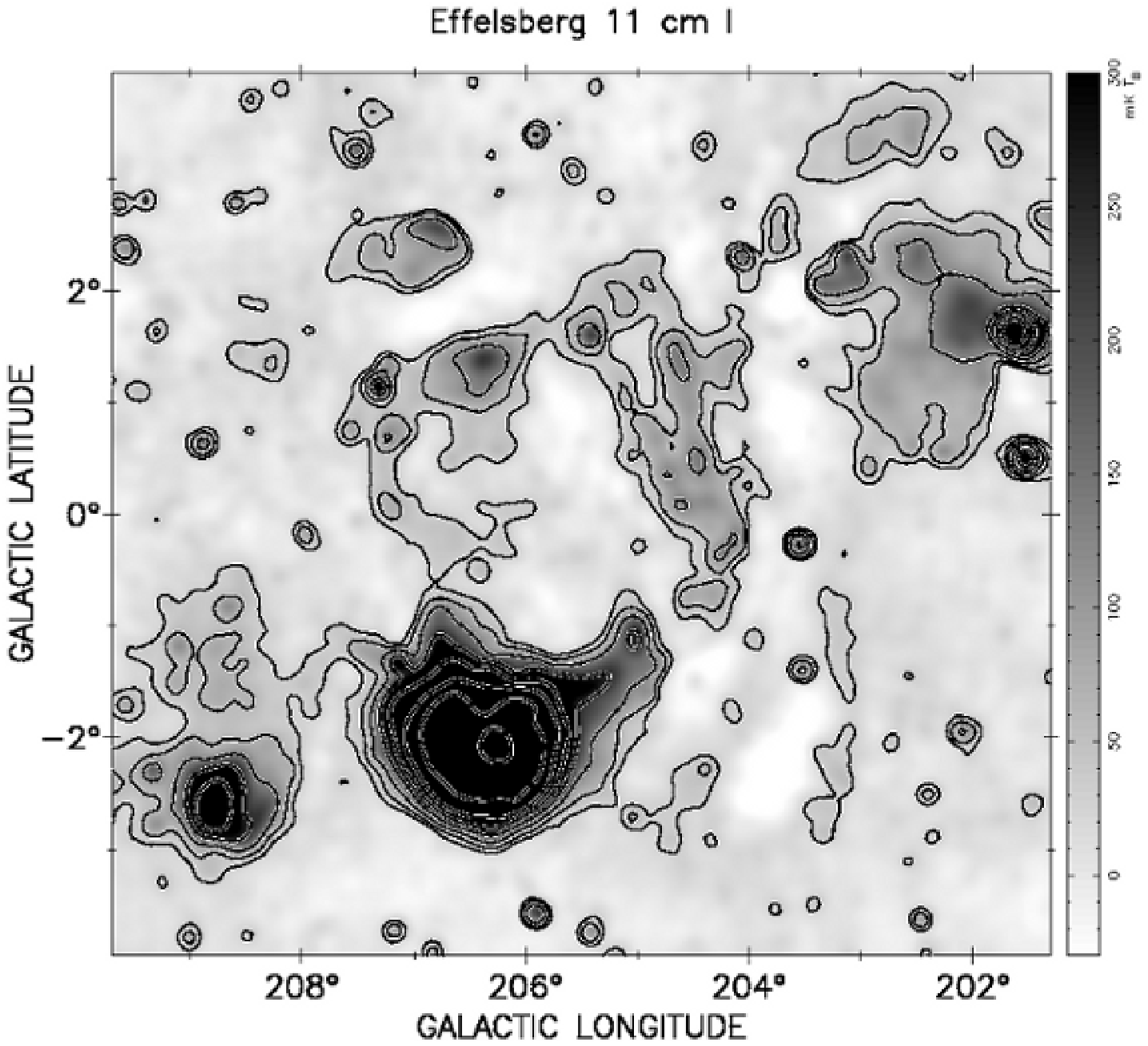}
\caption{The Effelsberg 21~cm and 11~cm continuum images of the Monoceros SNR,
    with angular resolutions of 9.4$\arcmin$ and 4.4$\arcmin$ respectively.
    Contours at 21~cm start from 100~mK~T$_{\rm B}$ and increase by
    2$^{n}\times$80~mK~T$_{\rm B}$ (n=0,1,2,3,$\cdots$). 
    Contours at 11~cm are drawn according to 2$^{n}\times$30~mK~T$_{\rm B}$, with n=0,1,2,3,$\cdots$.
}
\label{cont}
\end{center}
\end{figure}

\begin{figure}[!hbt]
\begin{center}
\includegraphics[angle=-90,width=0.48\textwidth]{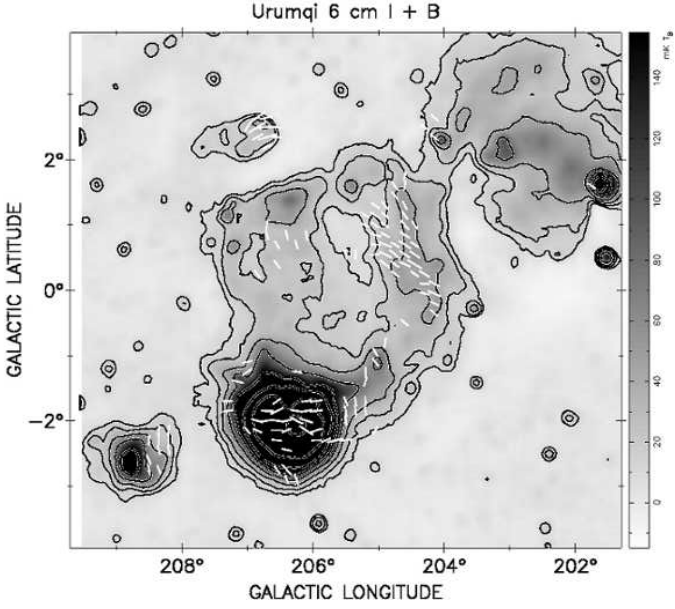}
\caption{The Urumqi 6~cm total intensity map of the Monoceros SNR at an angular resolution
    of 9.5$\arcmin$. Polarization vectors in the B-field direction (E+90$\degr$) are overlaid.
    The length of the bars is proportional to PI, with a polarized intensity of
    1~mK~T$_{\rm B}$ corresponding to a bar-length of $2\arcmin$. 
    Total intensity contours are drawn according to 2$^{n}\times$10~mK~T$_{\rm B}$, 
    with n=0,1,2,3,$\cdots$. 
}
\label{6cm}
\end{center}
\end{figure}

Figure~\ref{cont} shows the Effelsberg 21~cm and 11~cm total intensity images of the Monoceros
SNR. The Urumqi 6~cm continuum emission with corresponding polarization intensity (E-vectors)
overlaid are presented in Fig.~\ref{6cm}.  Filamentary shell structures are clearly seen in the
northern, western and southern regions with diffuse emission in the eastern portion. 
The HII region 0641+06 ($l$=206\fdg35, $b$=1\fdg35) was identified by~\citet{ghs82}. 
There is also a small emission ridge ($l=207\fdg6, b=-1\fdg3$) projecting from the Rosette Nebula
towards the east. In Sects. 3 and 5, we propose this structure to be part of the Monoceros
southern shell.

\subsection{HI data}
The HI data sets come from the recently released Galactic Arecibo L-Band Feed Array HI survey
(GALFA-HI), described in detail by~\citet{phd11}. The angular resolution of the survey is about
4$\arcmin$.  It covers a wide velocity range from $-$700~km~s$^{-1}$ to +700~km~s$^{-1}$ with a
velocity resolution of 0.18~km~s$^{-1}$. Typical noise levels are 80~mK in an integrated
1~km~s$^{-1}$ channel.

\section{Spectral index analysis}
\subsection{Subtracting the thermal radio continuum emission}
\begin{figure}[!hbt]
\begin{center}
\includegraphics[angle=-90,width=0.48\textwidth]{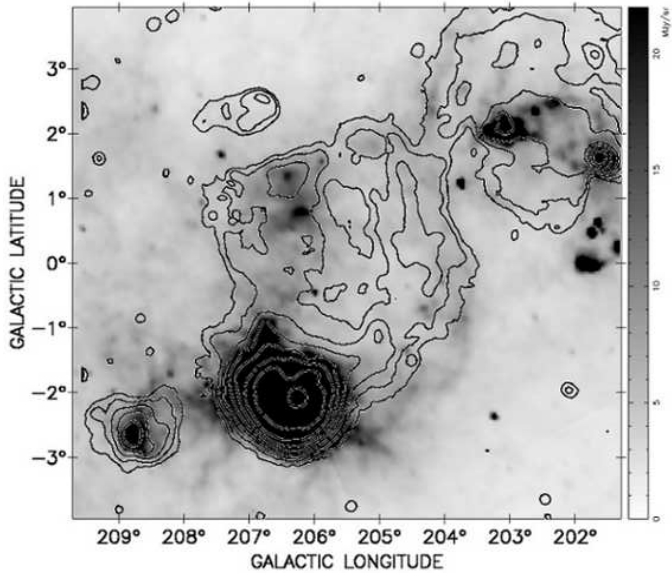}
\caption{IRIS 60 $\mu$m emission of the same area as in the preceding radio
    continuum images. Resolution is 2$\arcmin$. Contours of 6~cm total intensity
    are overlaid having the same levels as for Fig.~\ref{6cm}.
}
\label{dust}
\end{center}
\end{figure}

Thermal emission is well traced by the infrared (IR) emission from the dust. We present the IRIS
60 $\mu$m image of the Monoceros complex region in Figure~\ref{dust}.
The IR observations are the high-resolution reprocessed IRAS image produced at the Infrared
Processing and Analysis Center (IPAC)~\citep{ml05}. Besides the strong thermal emission from the
Rosette Nebula, there is a thermal source ($l$=206\fdg25, $b$=0\fdg75) within the SNR region and
HII region 0641+06 within the shell.


We tried to separate and subtract the aforementioned thermal sources using the correlation
between the radio brightness temperature and its equivalent dust flux at 60~$\mu$m
(e.g.~\citet{boh89,frs87}).  This correlation generally exists for diffuse thermal emission and
has been used to remove the thermal background emission from SNRs.

\begin{figure}[!hbt]
\begin{center}
\includegraphics[angle=-90,width=0.36\textwidth]{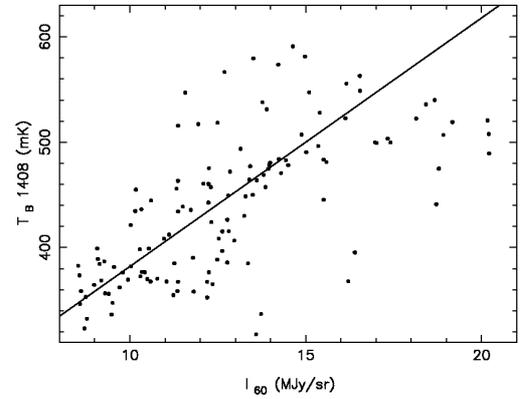}
\caption{Plot of the brightness temperature $T_{\rm B,21}$ against the brightness $I_{60}$.  
    	The best-fit straight line corresponds to a slope of 23.6~mK~MJy$^{-1}$~sr.
}
\label{slope}
\end{center}
\end{figure}

For diffuse HII regions,~\citet{boh89} obtained the relation $T_{\rm B,11}=(6.4\pm1.7)\times I_{60}$, where $T_{\rm B,11}$ is in mK and $I_{60}$ in MJy~sr$^{-1}$. Assuming a temperature
spectral index $\beta=-2.10$, the corresponding relation at 21~cm becomes
$T_{\rm B,21}=24.6I_{60}$. We made a TT-plot between 21~cm and 6~cm for the thermal source
region.  A temperature spectral index $\beta=-2.10\pm0.22$ confirms the thermal nature of the
emission.  We convolved the 21~cm and 60~$\mu$m images to the same 10$\arcmin$ resolution, and
plotted the brightness at 60~$\mu$m ($I_{60}$) against that at 21~cm ($T_{\rm B,21}$). The
resulting correlation is shown in Fig.~\ref{slope}, from which we find a slope of 
23.6~mK~MJy$^{-1}$~sr, comparable with the theoretical value. We were then able to
estimate the theoretical thermal emission and simply subtract it from the raw images. 
The thermal emission at 11~cm and 6~cm are simply interpolated using a spectral index of
$\beta=-2.10$.

We checked the similar relation for the Rosette Nebula. Both the Effelsberg 21~cm total intensity
data and the IRIS 60~$\mu$m data were convolved to 10$\arcmin$ and scaled to units of
Jy/beam. The ratio $T_{\rm B,21}/I_{60}$ shows large scatter in the periphery, with an average
value of about 90~mK~MJy$^{-1}$~sr in the central region. In regions where the two objects
overlap each other, it is difficult to separate the nonthermal emission from the SNR and the
thermal emission from the Rosette Nebula. Therefore we directly exclude the Rosette Nebula emission by setting a boundary in between to obtain the emission from the SNR alone.

\subsection{Integrated flux densities and TT-plots}
In order to estimate integrated flux densities and derive an accurate spectral
index for the remnant, we removed five bright point-like sources from the 21, 11
and 6~cm total intensity maps within the area of the Monoceros SNR by Gaussian fitting.

\begin{figure}[!hbt]
\begin{center}
\includegraphics[angle=-90,width=0.4\textwidth]{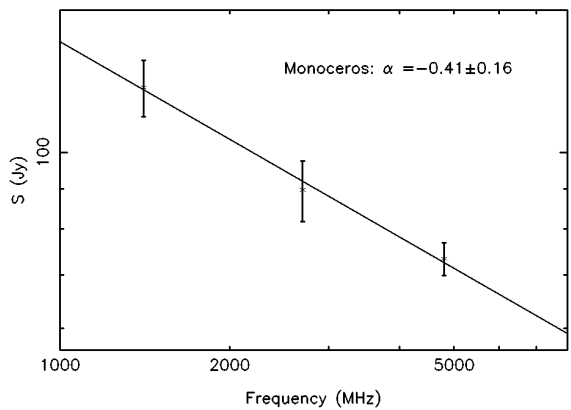}
\caption{Spectral index from integrated flux densities of the Monoceros SNR.
}
\label{spect}
\end{center}
\end{figure}


We used the method of temperature-versus-temperature plots (TT-plots)~\citep{tpk62}
to adjust the base-levels for the entire SNR area at two wavelengths. Both maps were convolved
to a common angular resolution of $10\arcmin$. Corresponding to the TT-plot results for the
entire shell region, we found a constant offset of 3~mK~T$_{\rm B}$ at 6\ cm, 55~mK~T$_{\rm B}$
at 11~cm and about 35~mK~T$_{\rm B}$ at 21~cm.

The integrated flux densities are obtained by setting polygons just outside the periphery of the
SNR and integrating the enclosed emission. The contribution of the Rosette Nebula was excluded by
setting the highest brightness temperature in the shell as its boundary. From variations
outside the SNR, we estimated a remaining uncertainty in setting the base-level of
9.9~mK~T$_{\mathrm B}$ at 21~cm, 7.9~mK~T$_{\mathrm B}$ at 11~cm, and 3.5~mK~T$_{\mathrm B}$ at
6~cm. We obtained a flux density of 120.8$\pm$9.9~Jy for the Monoceros SNR at 21~cm,
89.7$\pm$7.9~Jy at 11~cm and 73.4$\pm$3.5~Jy at 6~cm. The linear fit to these integrated flux
densities for the Monoceros SNR is shown in Fig.~\ref{spect}, 
yielding a spectral index of $\alpha = -0.41\pm$0.16.

\begin{figure*}[!hbt]
\begin{center}
\includegraphics[angle=-90,width=0.3\textwidth]{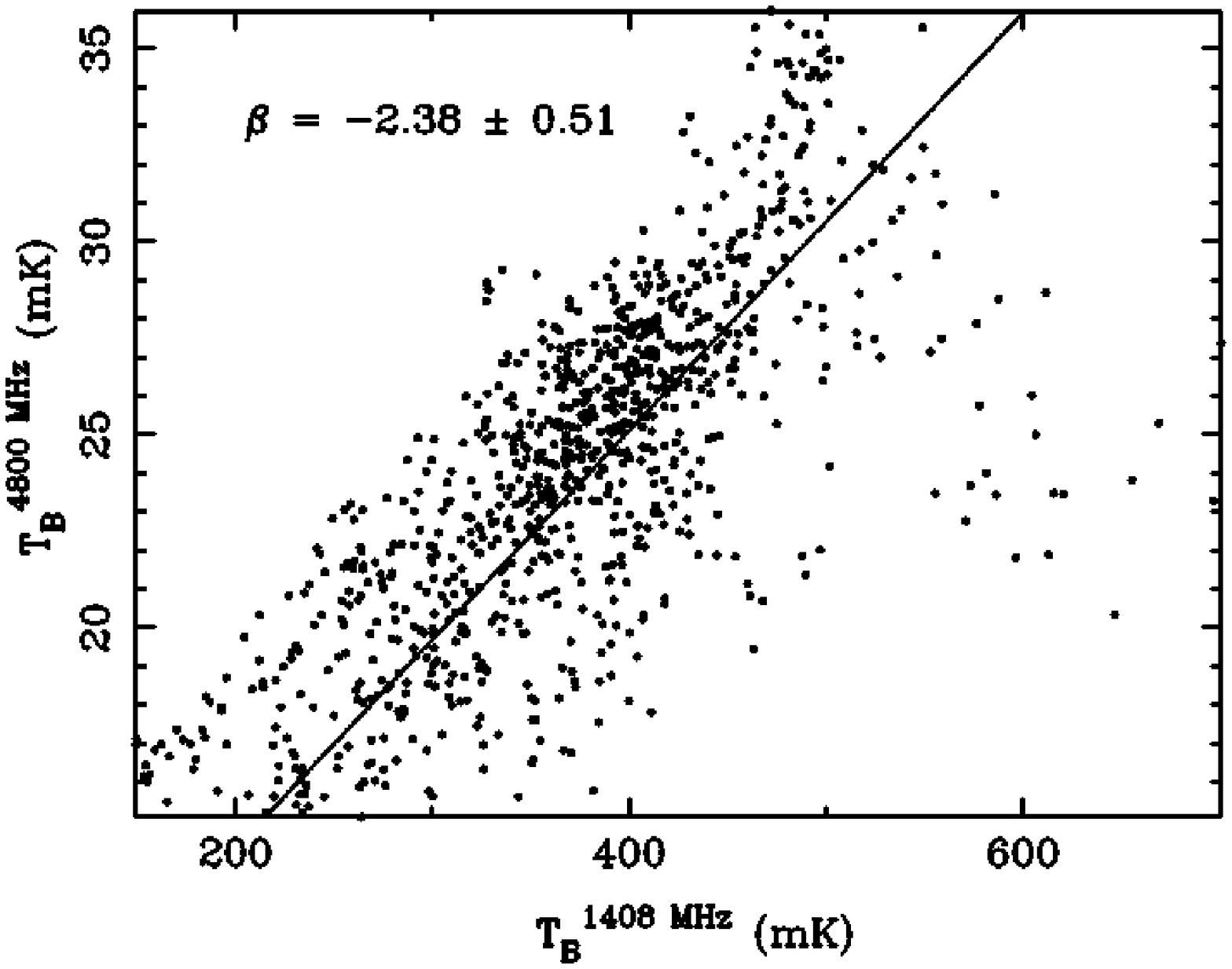}
\includegraphics[angle=-90,width=0.3\textwidth]{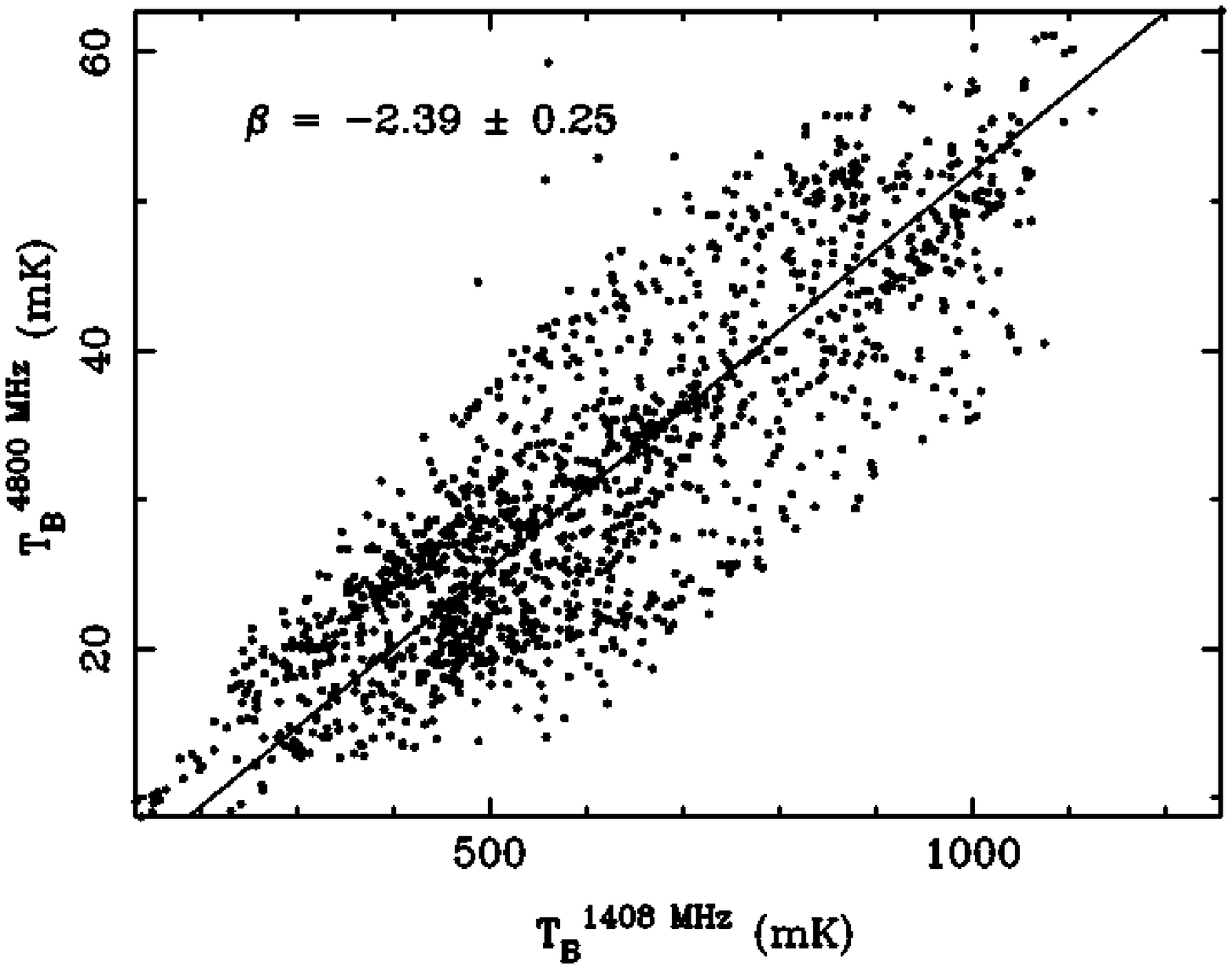}
\includegraphics[angle=-90,width=0.3\textwidth]{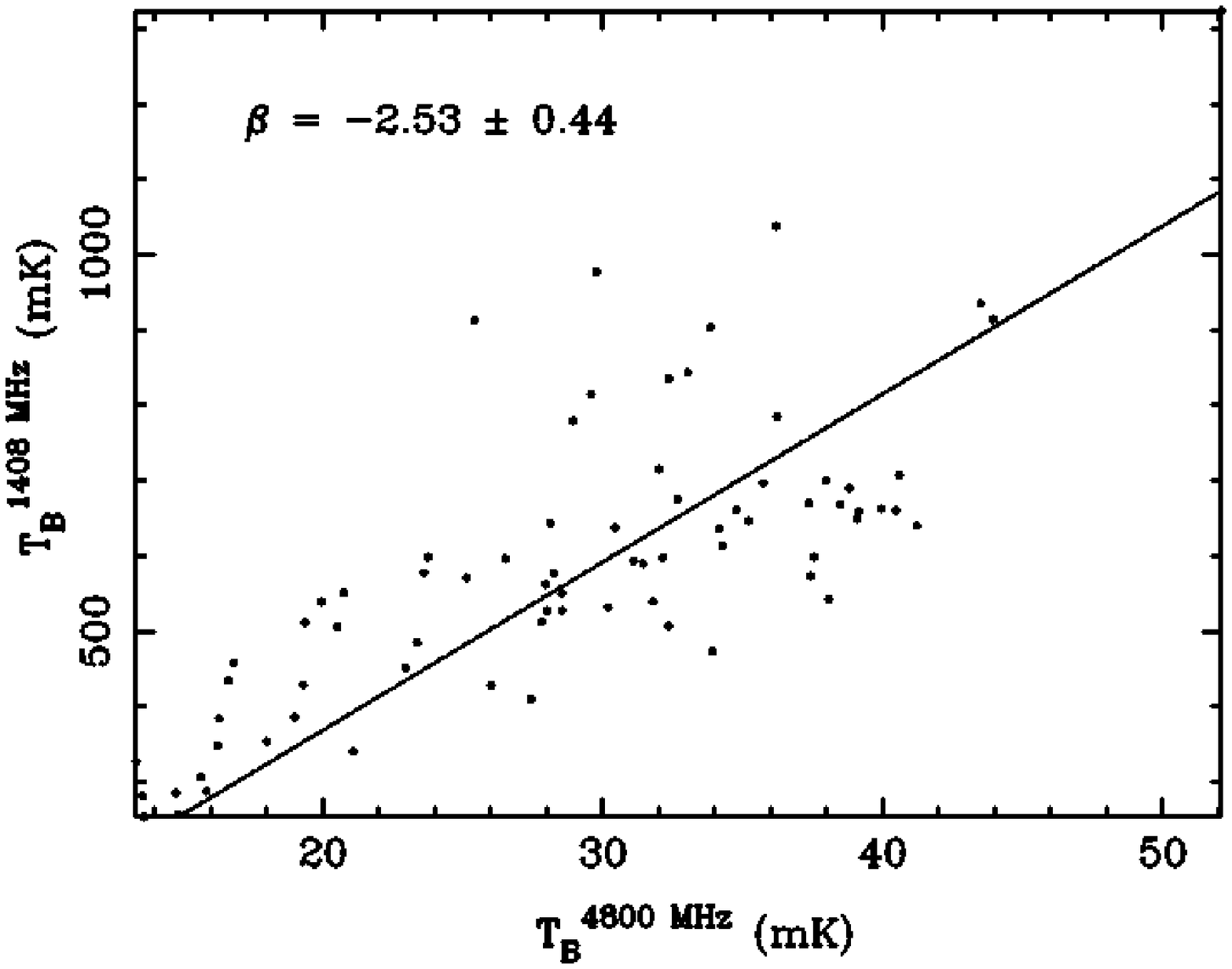}
\caption{TT-plots between Urumqi 6~cm data and the Effelsberg 21~cm data (as examples)
     for the eastern shell, northwest filamentary shell and newly identified
     southern branch shell of the Monoceros SNR.
}
\label{ttplot}
\end{center}
\end{figure*}

The TT-plot method is unaffected by the uncertainty of the background levels in the maps. 
We also applied it to an investigation of the spectral index for different sections of the
Monoceros SNR shell. The temperature spectral index $\beta~(\beta=\alpha-2)$ found from fitting
the slope of 21~cm and 6~cm pairs of different shell regions are shown in Fig.~\ref{ttplot}.
The error of $\beta$ for the diffuse eastern shell is relatively large, probably due to its
low brightness and confusion with weak unresolved background sources which could not be
subtracted. We also investigated the southern shell region which is mixed with the Rosette
Nebula. Beside the old spherical shell region, there is the new shell-like feature
protruding out towards the east (see Section 2.1). The index of the new shell-like feature
is about $\alpha=-0.53\pm0.44$. The large uncertainty is due to the weak emission, plus the
residuals after a point-source has been subtracted. However, the spectral index distribution in
this region in  Fig. 4 of~\citet{ghs82}, as well as Fig. 7 in Sect. 3.3, are compatible with the
feature being nonthermal. It seems that this feature is likely to be a part of the
Monoceros SNR.

\subsection{Spectral index map}
\begin{figure}[!hbt]
\begin{center}
\includegraphics[angle=-90,width=0.48\textwidth]{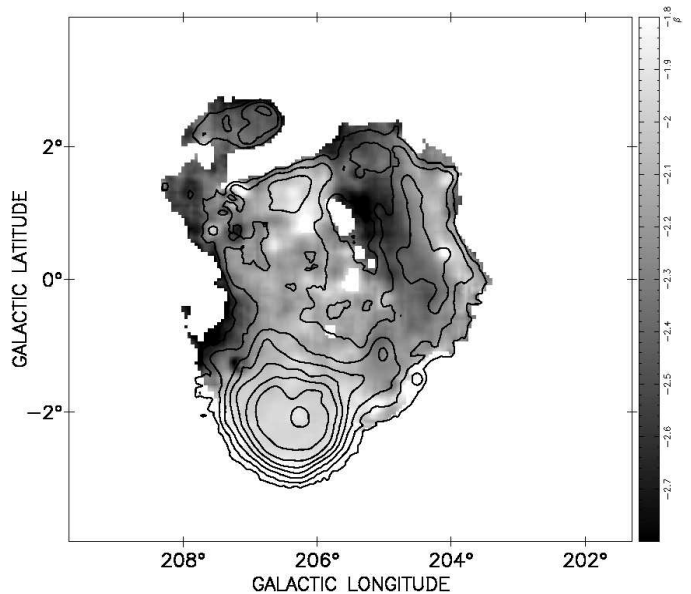}
\caption{Temperature spectral index map of the Monoceros SNR from 21/11/6~cm maps.
    Regions where the random error exceeds 0.06 have been blanked out.
    Contours of 6~cm total intensity with the same levels as Fig.~\ref{6cm} are overlaid.
}
\label{index}
\end{center}
\end{figure}

We calculated the spectral index for each pixel of the Monoceros SNR from the zero-level corrected
images at the three frequencies, by linearly fitting for intensities versus frequencies
using a logarithmic scale. In order to achieve reasonable spectral indices without
significant influence from noise or local distortions, a lower intensity limit of
180~mK~T$_{\rm B}$, 30~mK~T$_{\rm B}$ and 6~mK~T$_{\rm B}$ are set for the 21, 11 and 6~cm
maps, respectively. The regions outside the remnant are blanked out. We display the
temperature spectral index map from 21~cm, 11~cm and 6~cm in Fig.~\ref{index}. Possible
remaining variations of the base-levels at 11~cm or 6~cm yield an average systematic uncertainty
of the spectral indices of $\Delta\alpha\sim0.2$ with errors being larger in regions of
weak emission.

The Rosette Nebula shows a flat brightness spectral index, as expected for optically thin
thermal emission from an HII region. The spectral index of the SNR G206.9+2.3 shows fairly uniform
distribution and has an averaged index of $\beta=-2.46$, similar to the value found for its
integrated flux densities by~\citet{ghr11}. Spectral index values are generally about $-0.4$
in the Monoceros SNR shell region. The new shell-like feature in the southern shell has an index
of about $-0.5$. Little significant spectral variation is traced in the overlap region
with the Rosette Nebula. The spectral index gets steeper in the inner region so that the inner
western filament at $l=205\fdg5, b=1\fdg0$ has a steeper spectral index of about $-0.6$.
The value is generally consistent with that in the lower frequency spectral index map between
178 and 2700~MHz~\citep{ghs82}.

\citet{ghs82} excluded the possibility that the difference comes from a mixture of thermal
emission for the more northerly rim of NGC 2264 using a positional argument and by
comparison with the lower frequency index. The new higher frequency 6~cm data further
excludes this possibility.  The relativistic electrons in the inner western ridge probably have
a different energy spectrum as proposed by~\citet{ghs82}. It is now generally accepted that
relativistic electrons responsible for the synchrotron emission are produced by the Fermi
acceleration mechanism~\citep{b78a,b78b}.  This process predicts a spectral index given by
$-\beta=2+(M^{2}+3)/[2(M^{2}-1)]$, where $M$ is the Mach number of the shock. For a very strong
shock, $M$ is close to 3, corresponding to an average value of $\beta=-2.75$. A weaker shock
produces a steeper spectrum. The shock in the inner western filament is possibly weaker with a
decreased compression ratio and results in a steeper spectrum. In addition, a steeper spectrum
could also have originated from high-energy electrons due to synchrotron aging, when propagating
in a weak magnetic field region.  However, the strength of the magnetic field in both western
ridges appears to be similar from the vectors in the 6~cm map, both from the perpendicular
component B$_{\perp}$ and the amounts of Faraday rotation. No indication of a spectral break
is found in the frequency range from 178~MHz to 4.8~GHz, which means that any spectral break due
to synchrotron aging should be at higher frequencies. Higher frequency data are needed to
distinguish between these two explanations.

\section{Polarization analysis}
\subsection{The linear polarization properties}
The 6~cm polarization map is shown in Fig.~\ref{6cm}. Prominent polarized emission is mainly
displayed towards the western shell of the SNR. The observed magnetic vectors are nearly
distributed along the ridges of the two branches. This indicates a well ordered magnetic
field in the shell, as expected for an evolved SNR~\citep{fr04}. The polarization percentage of
the inner filament is about 20\% at 6~cm, and less at about 10\% in the outer filament.
Except for the enhanced polarization along the SNR's western shell, polarized emission
in other shell regions is very weak. The inner polarization patch might originate from
polarized diffuse interstellar emission possibly mixed with polarized emission from the SNR of
about similar strength. Strongly polarized emission is also detected from the SNR G206.9+2.3.

The large-scale emission component is missing in the 6~cm polarization map, due to the zero-level
setting during observation and data analysis. This is known to limit the interpretation of
polarization structures caused by Faraday effects in the interstellar medium~\citep{r06} except
in strongly polarized regions. We checked the polarization map with large-scale polarized
emission added, which is extrapolated from the WMAP K-band data using an appropriate spectral
index (see Fig. 6 in ~\citet{grh10}). Strongly polarized emission from the diffuse interstellar medium appears near the Galactic plane. The polarized signal from the western shell
of the Monoceros SNR is sufficiently strong that it remains almost unchanged by
adding the large-scale emission component.

\begin{figure}[!hbt]
\begin{center}
\includegraphics[angle=-90,width=0.45\textwidth]{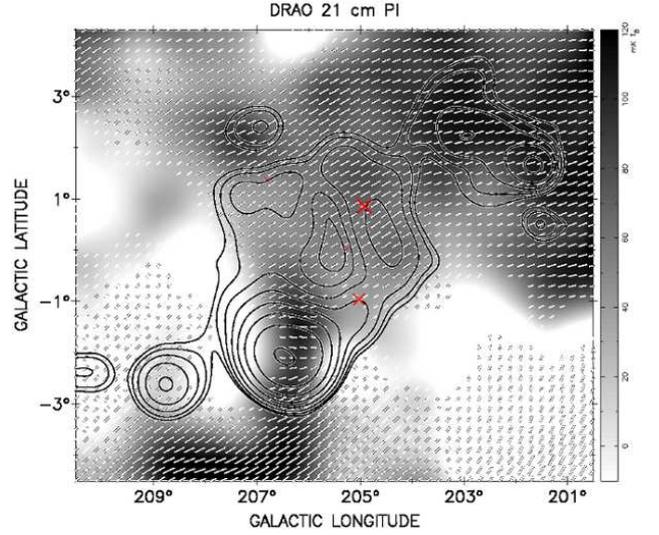}
\caption{DRAO 21 cm polarized intensity map of the Monoceros SNR at an angular resolution
    of 36$\arcmin$. Polarization vectors in the B-field direction (E+90$\degr$) are overlaid.
    The length of the vectors is proportional to PI. A polarized intensity
    of 1~mK~T$_{\rm B}$ corresponds to a length of 0.05$\arcmin$. Contours show total
    intensities starting at 200~mK~T$_{\rm B}$ and increasing by
    2$^{n}\times$60~mK~T$_{\rm B}$ (n=0,1,2,3,$\cdots$). The positions of RM sources in
    ~\citet{tss09} are indicated by crosses, with lengths proportional to the RM values.
}
\label{21cmpi}
\end{center}
\end{figure}

We extracted the 21~cm polarized intensity map of the same region (Fig.~\ref{21cmpi})
from the polarization survey made with the DRAO 26-m telescope~\citep{wlr06}. This map includes
the large-scale polarized emission component at an absolute zero level with an angular resolution
of 36$\arcmin$. We found a large diffuse interstellar polarization patch covering the Monoceros
SNR region, with enhanced polarized emission coinciding with the western filamentary
shell region. The polarization angles in the western shell lie nearly perpendicular to the
shell. Polarized emission from a larger distance might has been depolarized towards the Rosette
Nebula, where weakly polarized emission is detected at 21~cm.

Compared to the western shell, the low polarization percentage of the eastern shell
indicates a more irregular magnetic field. The background polarization there is depolarized by
small-scale magneto-ionic fluctuations in the shell and/or polarized emission from the SNR shell
with a different orientation. In the southern shell where interaction has occurred, as discussed
in Sect. 5.3, the high thermal electrons densities in the Rosette Nebula~\citep{o86} and
their fluctuations could be sufficiently strong to cause a heavy depolarization of
the background polarization.

\subsection{Rotation measures of the western shell}
Since the polarized emission from the western shell is strong and clearly stands out against the
background in the 21~cm polarization map, we were able to remove the diffuse Galactic polarized
emission component and calculate the rotation measure (RM) distribution using the
polarization angles from the 21 and 6~cm polarization maps. We subtracted an average value
of 81~mK~T$_{\rm B}$ from the 21~cm $U$ map and "twisted" the $Q$ map setting by values at
the four corners; (from upper left to lower left these values are 8, 80, 4, 8
~mK~T$_{\rm B}$). The 6~cm $U$ and $Q$ data were smoothed to an angular resolution of 36$\arcmin$
and we re-derived a map of polarization angles. We linearly fit the polarization angles at 21~cm
and 6~cm versus the square of the wavelength for each pixel from all maps. To gain a high signal
to noise ratio, pixels with brightness temperatures less than 3$\times \sigma_{PI}$ were not
included.

In principle, we require polarization maps for at least three frequencies to obtain
unambiguous RMs. RMs calculated at 21~cm and 6~cm have an ambiguity of
$\pm n\times 77$~rad~m$^{-2}$ (n=0,$\pm$1, $\pm2$,...). The average minimum RM (n=0) is about
+30~rad~m$^{-2}$ in the western filamentary region. Larger RMs are also possible in this case.
We checked the RMs of extragalactic sources towards this region from the catalog
by~\citet{tss09}. There is an extragalactic source ($l$=204$\fdg$92, $b$=0$\fdg$86) with an RM of
202.1$\pm6.6$~rad~m$^{-2}$ shinning through this region, and another source
($l$=205$\fdg$02, $b=-0\fdg$96) lies in the southeast shell with an RM of
152.7$\pm$5.4~rad~m$^{-2}$. Two sources within the Monoceros SNR region have RMs of
47.1$\pm$1.8~rad~m$^{-2}$ and 58.3$\pm$2.2~rad~m$^{-2}$ respectively (see Fig.~\ref{21cmpi}).
Though the RMs of the extragalactic sources represent RMs integrated through the total
interstellar medium, the RM values of about 109~rad~m$^{-2}$ (n=1) and 184~rad~m$^{-2}$ (n=2)
are more likely from the western shell, considering the distance of the Monoceros SNR of 1.6~kpc.
In Fig.~\ref{cont} we noticed that the polarization B vectors at 6~cm in the western shell
as shown are distributed almost tangential to the shell. If we accept deviations of about
20$\degr$ from an exactly tangential orientation, the RM is estimated to be about
96~rad~m$^{-2}$, near to the RM value for $n=1$ of 109~rad~m$^{-2}$.
Detailed investigation of the unambiguous RMs requires polarization observations at other
wavelengths.

\subsection{Strength of the magnetic field in the western shell}
We can estimate the magnetic field strength in the western shell by assuming energy
equipartition between the magnetic field and electrons and protons in the SNR-shell. We used the
revised equipartition formula by~\citet{bk05}, which uses the number density ratio $K$ instead of
the total energies of cosmic ray protons and electrons in classical equipartition estimates.
Assuming the synchrotron emission is concentrated in the SNR filamentary shell, the radio
emissivity $L_{\nu}/V$ averaged over the source's volume is replaced by the surface brightness
over the line-of-sight path length through the emitting region $I_{\nu}/l$. Therefore:
\begin{equation}
B_{eq}=\big\{ \frac{4\pi(-2\alpha+1)(K_{\rm o}+1)I_{\nu}E_{p}^{1+2\alpha}(\nu/2c_{1})^{-\alpha}}
{(-2\alpha-1)c_{2}(-\alpha)l}\big\}^{1/(-\alpha+3)}
\end{equation}
where, $I_{\nu}$ is the synchrotron intensity at frequency $\nu$ and $\alpha$ the synchrotron
spectral index of $-0.4$. $K_{\rm o}$ is a constant ratio of number densities of protons and
electrons in the energy range, which is assumed to be 1. $E_{p}$ is the proton rest energy.
$c_{1}$ and $c_{2}$ are constants (see the Appendix of~\citet{bk05}). We integrated the
flux density at 6~cm for the western shell section centered at $l=204\fdg6,~b=0\fdg5$
(($\delta l, \delta b)=2\fdg3\times0\fdg8$), and obtained a value of 11~Jy. The radio intensity
of the western shell is about $\rm I_{\nu}=0.2\times 10^{-18} erg~s^{-1} cm^{-2} Hz^{-1} sr^{-1}$. 
We estimate the thickness of the western filament from high angular resolution 11~cm data to be
$\sim13\arcmin$, about 10\% of the radius of the SNR. The thickness of the filament is about 6~pc,
and the total path through the interacting zone is $l\sim 49$~pc at a distance of 1.6~kpc. 
The estimated equipartition magnetic field $\rm B_{eq}$ is about $9.5~\mu$G, corresponding to a
factor of $\sim3-4$ enhancement of the mean ISM field of B$_{tot}$~2$-$3$\mu$G~\citep{hml06}.

\begin{figure*}[!htbp]
\includegraphics[width=0.33\textwidth, angle=0]{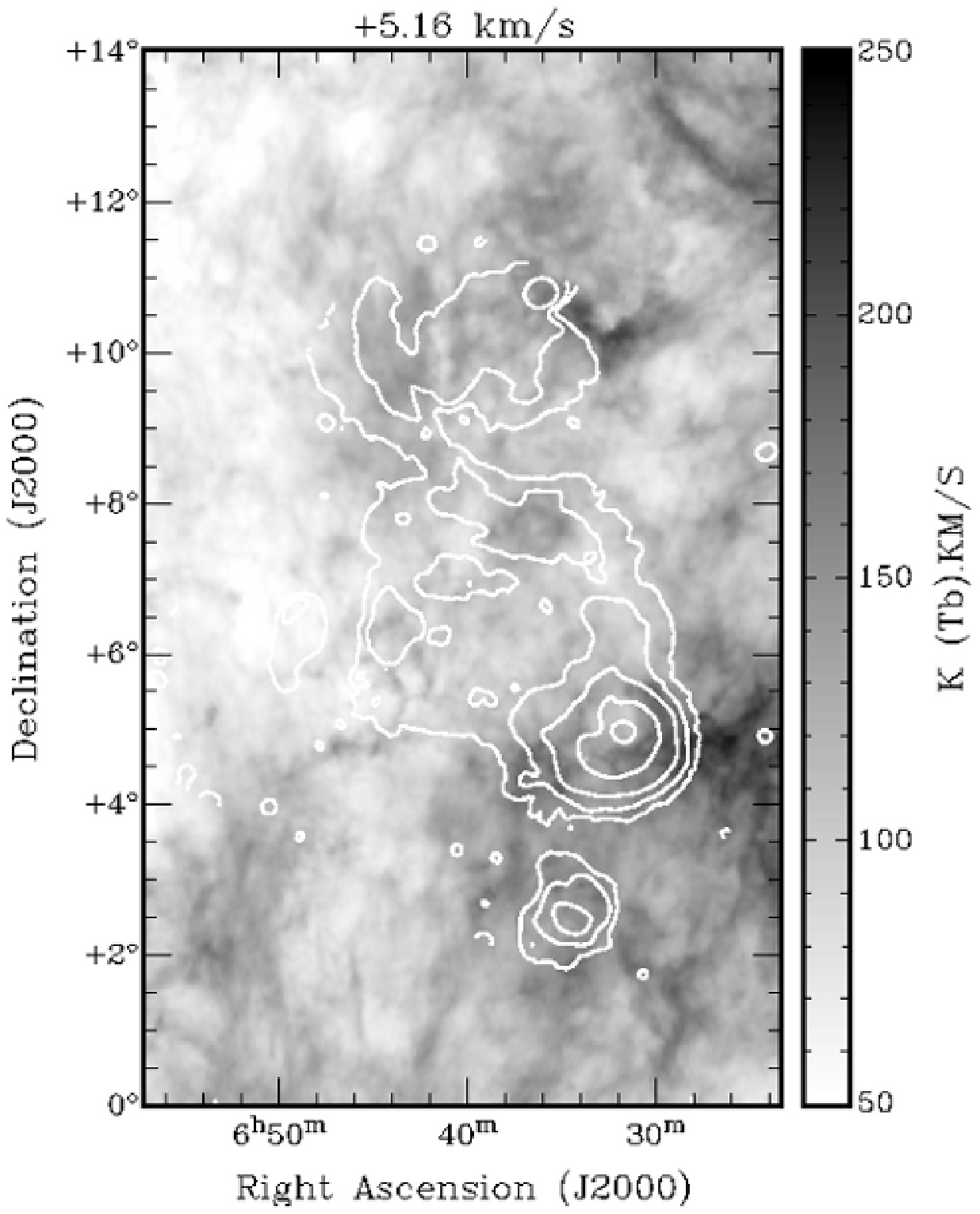}
\includegraphics[width=0.33\textwidth, angle=0]{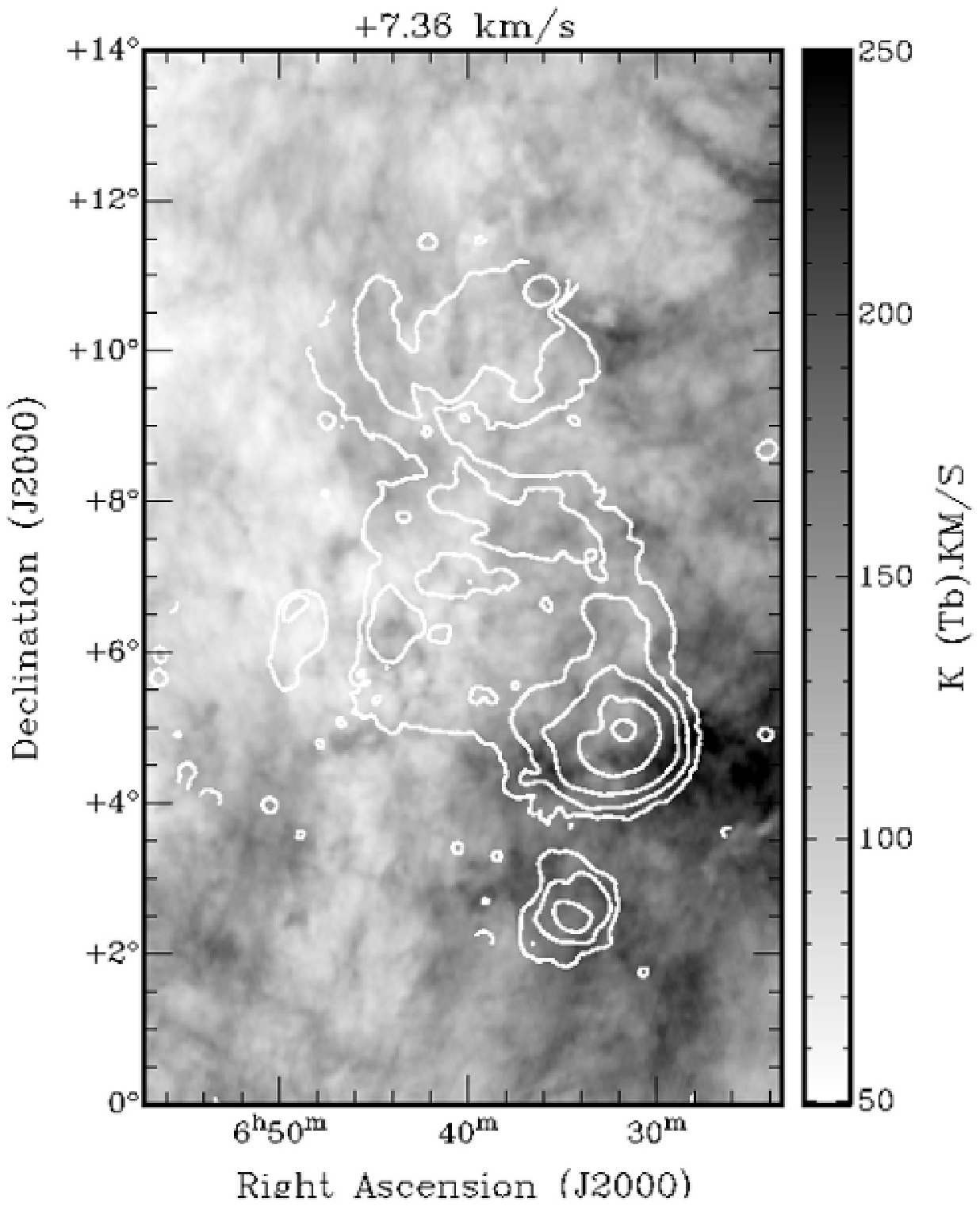}
\includegraphics[width=0.33\textwidth, angle=0]{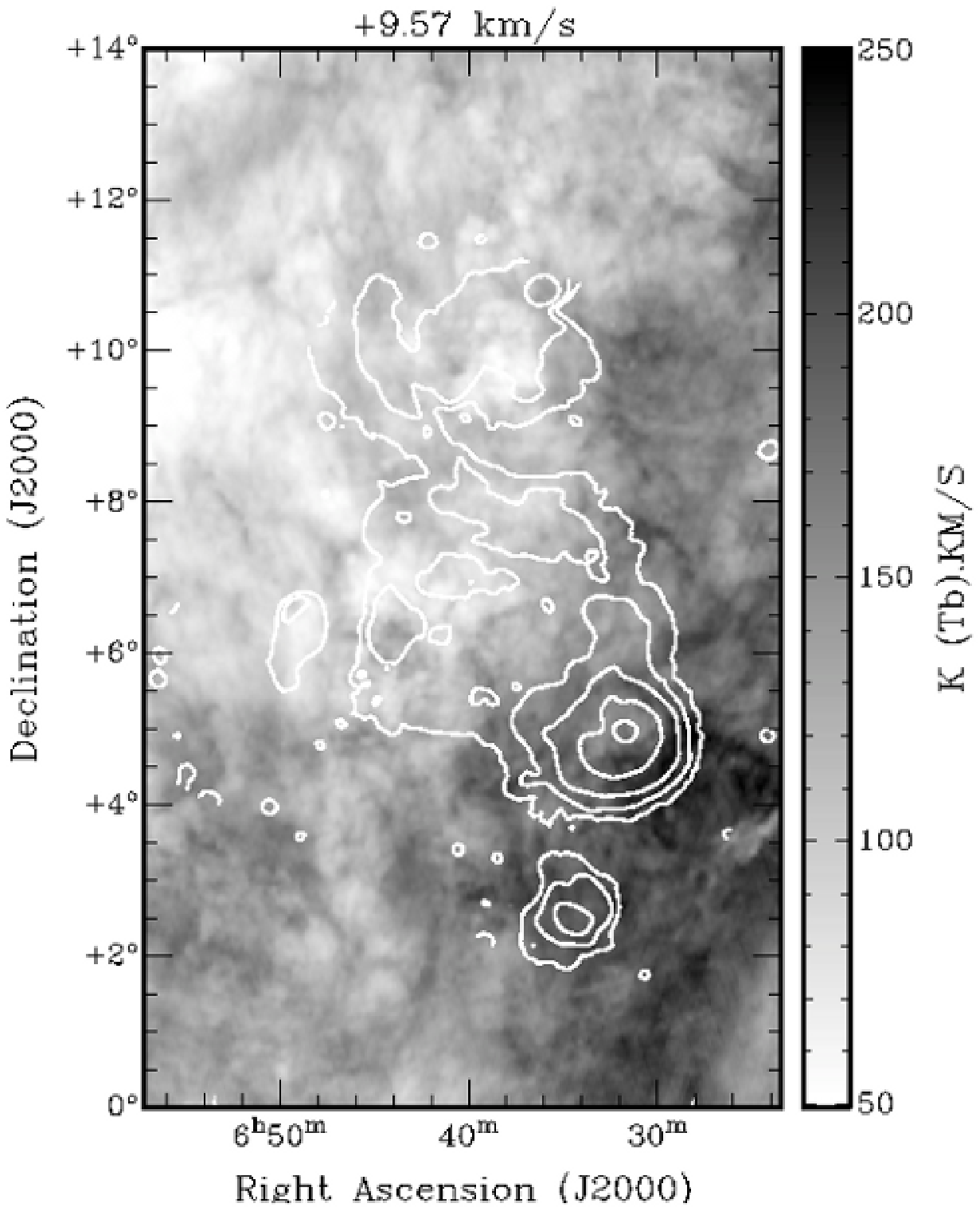}
\includegraphics[width=0.33\textwidth, angle=0]{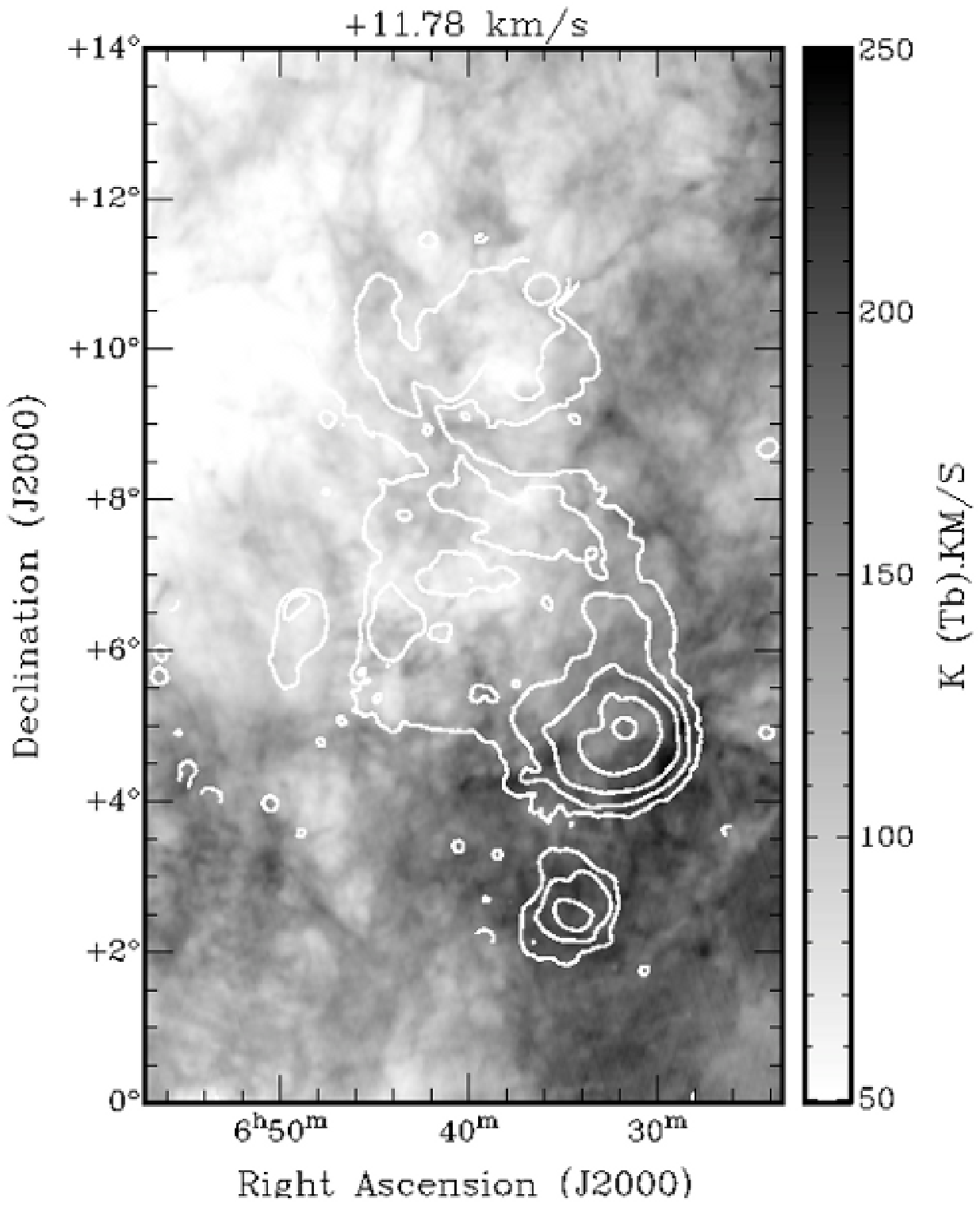}
\includegraphics[width=0.33\textwidth, angle=0]{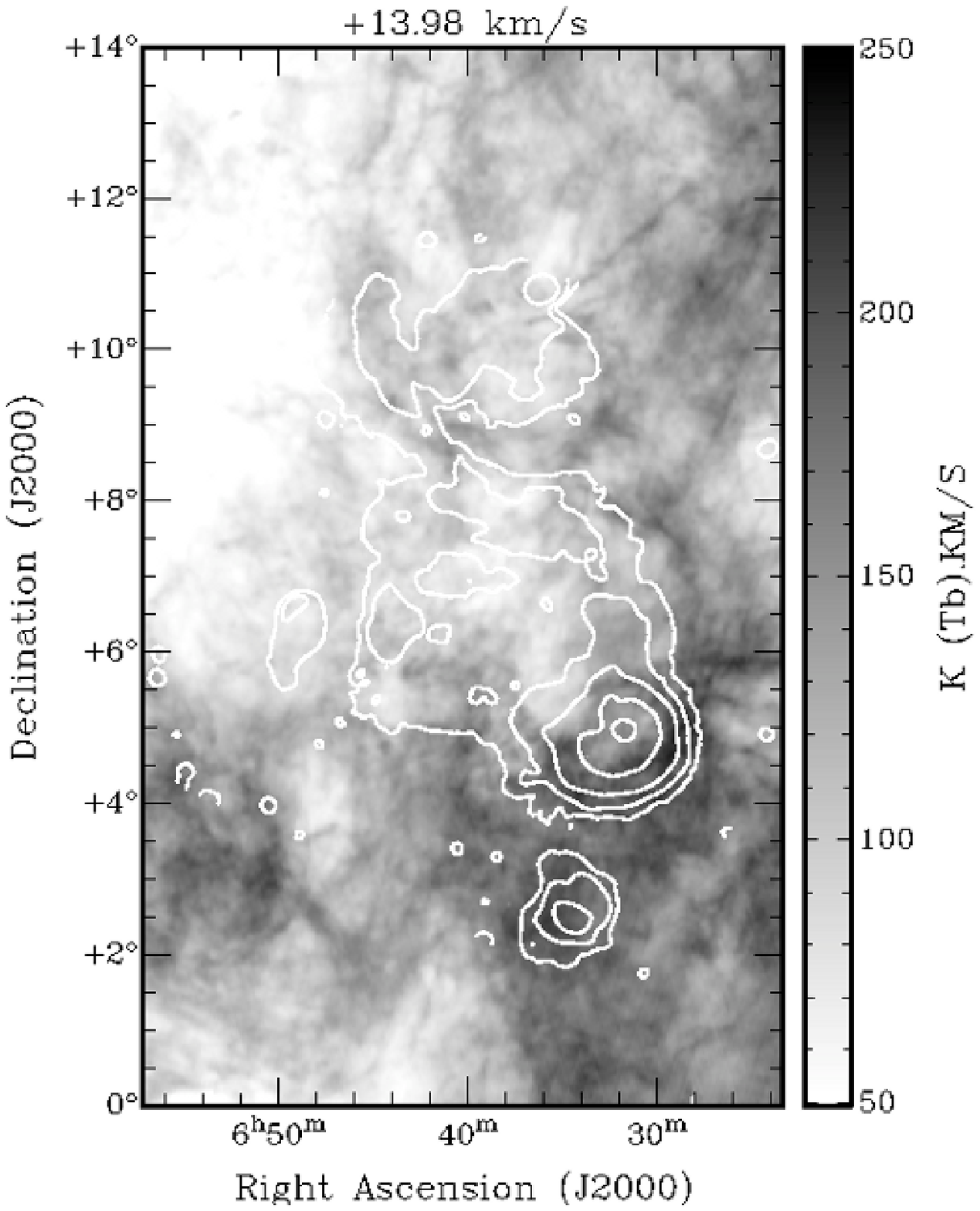}
\includegraphics[width=0.33\textwidth, angle=0]{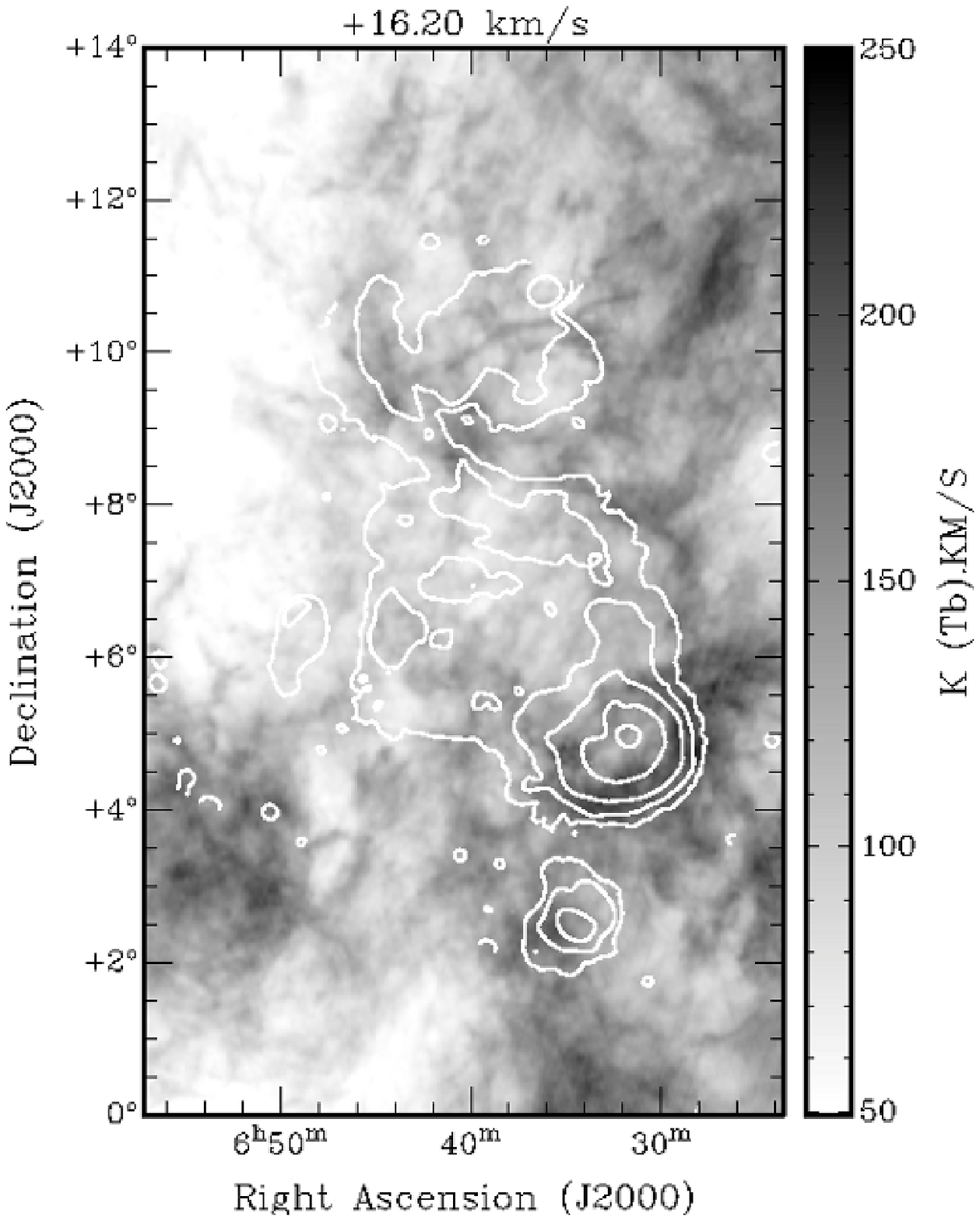}
\includegraphics[width=0.33\textwidth, angle=0]{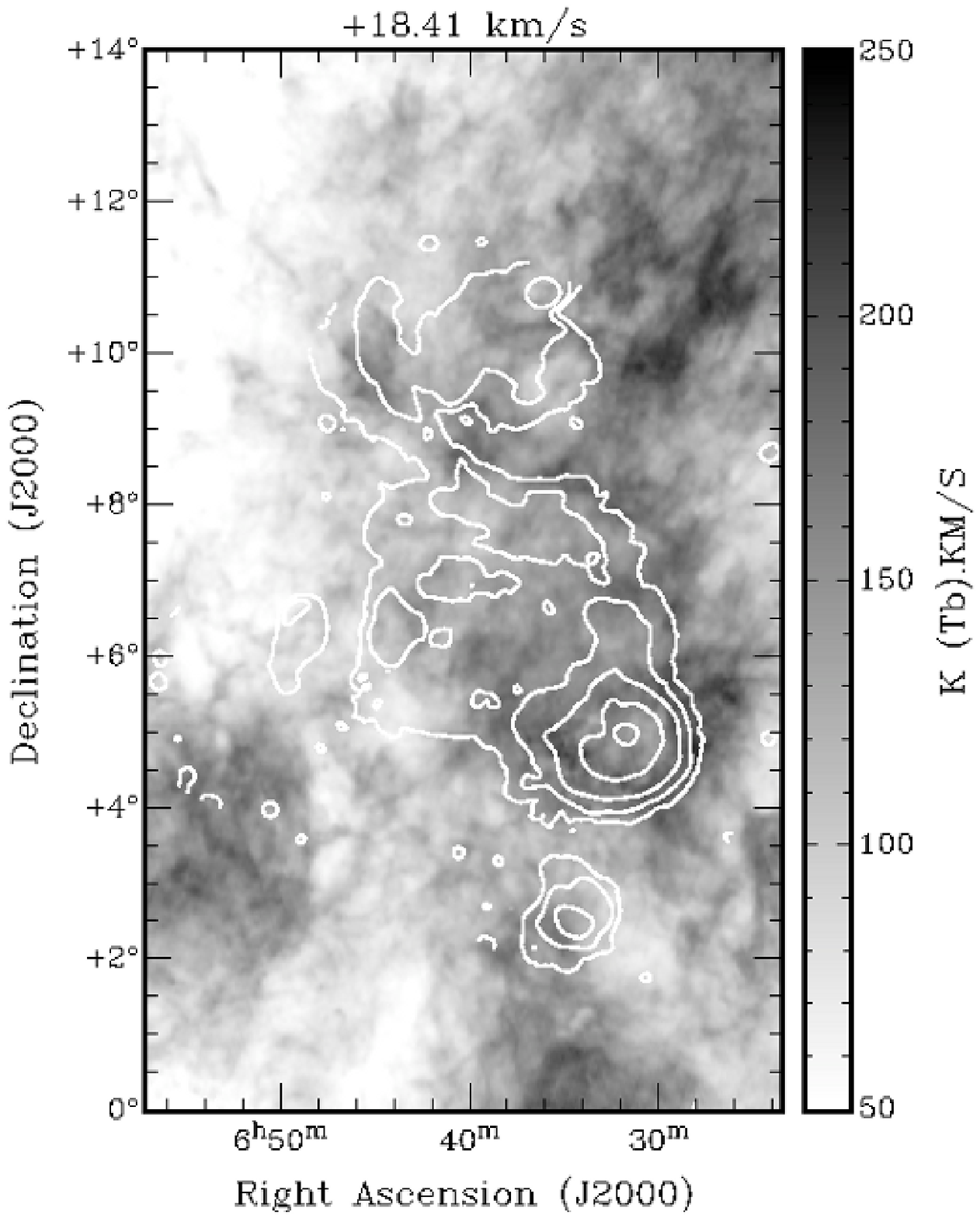}
\includegraphics[width=0.33\textwidth, angle=0]{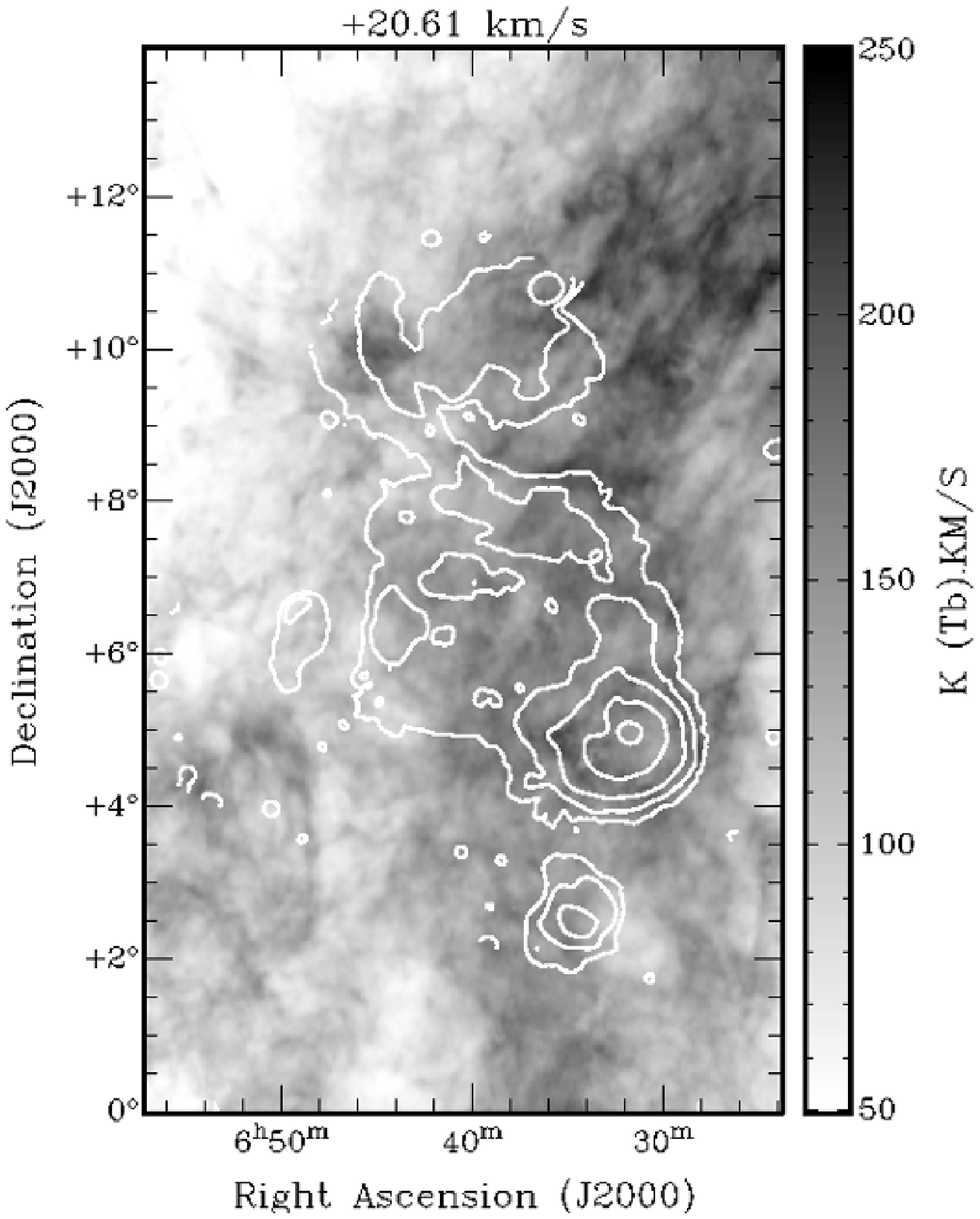}
\includegraphics[width=0.33\textwidth, angle=0]{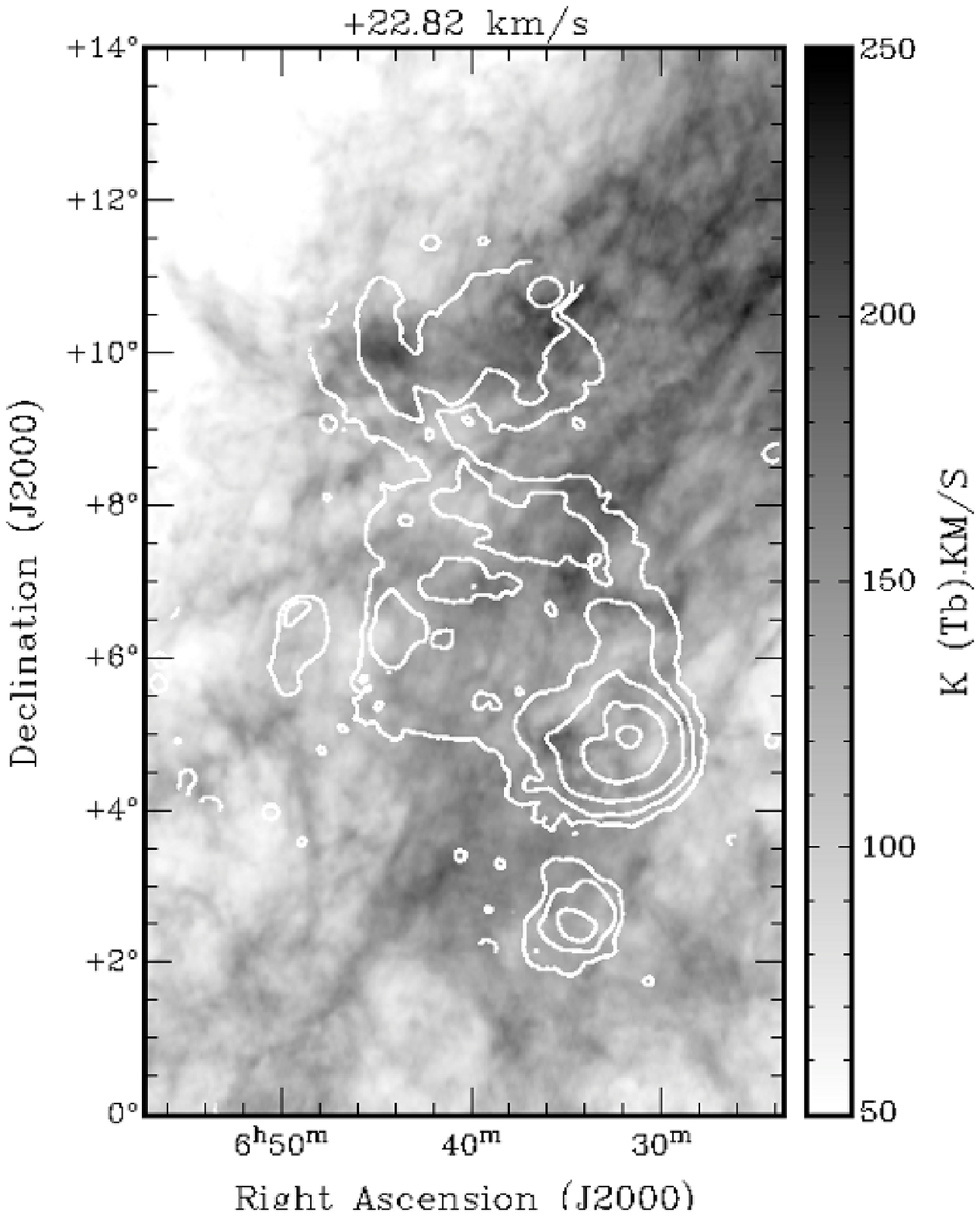}
    \caption{HI channel-maps in the area of the Monoceros SNR for the velocity range 
             from +4~km~s$^{-1}$ to +25~km~s$^{-1}$, averaged over 11 consecutive channels,
	     yielding a velocity resolution of $\sim$2~km~s$^{-1}$.
             The central LSR velocities are indicated above each panel.
             Total intensities of 6 cm continuum emission smoothed to $10\arcmin$ are
	     overlaid as contours.
}
    \label{channel}
\end{figure*}

\section{Neutral hydrogen morphology}
\subsection{HI associated with the Monoceros SNR}
The GALFA-HI survey covers a velocity range from $-$700~km s$^{-1}$ to +700~km s$^{-1}$ with a
velocity resolution of 0.18~km~s$^{-1}$. ~\citet{r66} and~\citet{r72} have shown that a
neutral hydrogen shell is located outside of the Monoceros optical filament with an HI cavity
inside. The GALFA-HI image has much better resolution and sensitivity and reveals many detailed
structures. 

The 0.8$-$1.6~kpc distance corresponds to a velocity range of $\rm +8-15~km~s^{-1}$ according
to the Galactic rotation model (\citep{fbs89}: $\rm R_0 = 8.5~kpc$ and
$\rm \Theta_0 = 220~km~s^{-1}$). We have inspected the entire data cube of this complex
region from the GALFA-HI survey in the velocity range between $\rm 0~km~s^{-1}$ and
$\rm +30~km~s^{-1}$, looking for structures morphologically related to the SNR.

Fig.~\ref{channel} displays the HI channel maps after averaging over 11 consecutive channels
($\sim$2~km~s$^{-1}$) within the velocity interval from +$4$ to $\rm +25~km~s^{-1}$. The central
velocity of each image is indicated at the top. Contours of 6\ cm total intensity (smoothed to
$10\arcmin$) are overlaid to indicate the Monoceros complex region. Towards this region, neutral
hydrogen filamentary structures caused by stellar winds from the OB associations of Mon OB1 and
Mon OB2 are visible across the entire region. The large-scale distribution of neutral hydrogen
gas shows a density gradient towards the Rosette Molecular Complex (RMC). The HI shell related
to the RMC lies between +5~km~s$^{-1}$ and +25~km~s$^{-1}$, with a lower emission towards the
southern shell of the Monoceros SNR, which is consistent with the result in~\citet{kb93}.

We have identified a partial neutral hydrogen shell surrounding the western SNR, which we believe
is associated with the SNR. This shell structure is distributed over velocities between
+7~km~s$^{-1}$ and +22~km~s$^{-1}$ with a thickness of $\sim$13$\arcmin$. It lies slightly
outside the spherical outline of the remnant, in particular in the southwest direction using
Galactic coordinates, and seems to have expanded ahead of the radio shell. This is similar
situation to the case of another likely old SNR, the North Polar Spur (Loop I)~\citep{bhs70}.
However, the HI shell is well associated with the 60~$\mu$m emission (Fig.~\ref{dust}), suggesting
that this shell could have been blown-out by strong stellar winds from the progenitor of the SNR
before the explosion. The 21~cm line emission is weak in the eastern SNR sector.
It seems that the SNR blast wave has expanded into a medium of lower density in this direction.
In the southern SNR region, the HI shell features are mixed with those of the Rosette Nebula.
However, a small section of possible shell structure is seen in the HI emission for the velocity
range +10 to 14~km~s$^{-1}$, projecting from the east of the Rosette Nebula.
This is prominent, being 0$\fdg$5 long with a thickness of 10$\arcmin$
in the +12~km~s$^{-1}$ channel map, and is well associated with the new shell-like 
radio continuum feature in the southern shell.

\begin{figure}[!hbt]
\begin{center}
\includegraphics[angle=0,width=0.5\textwidth]{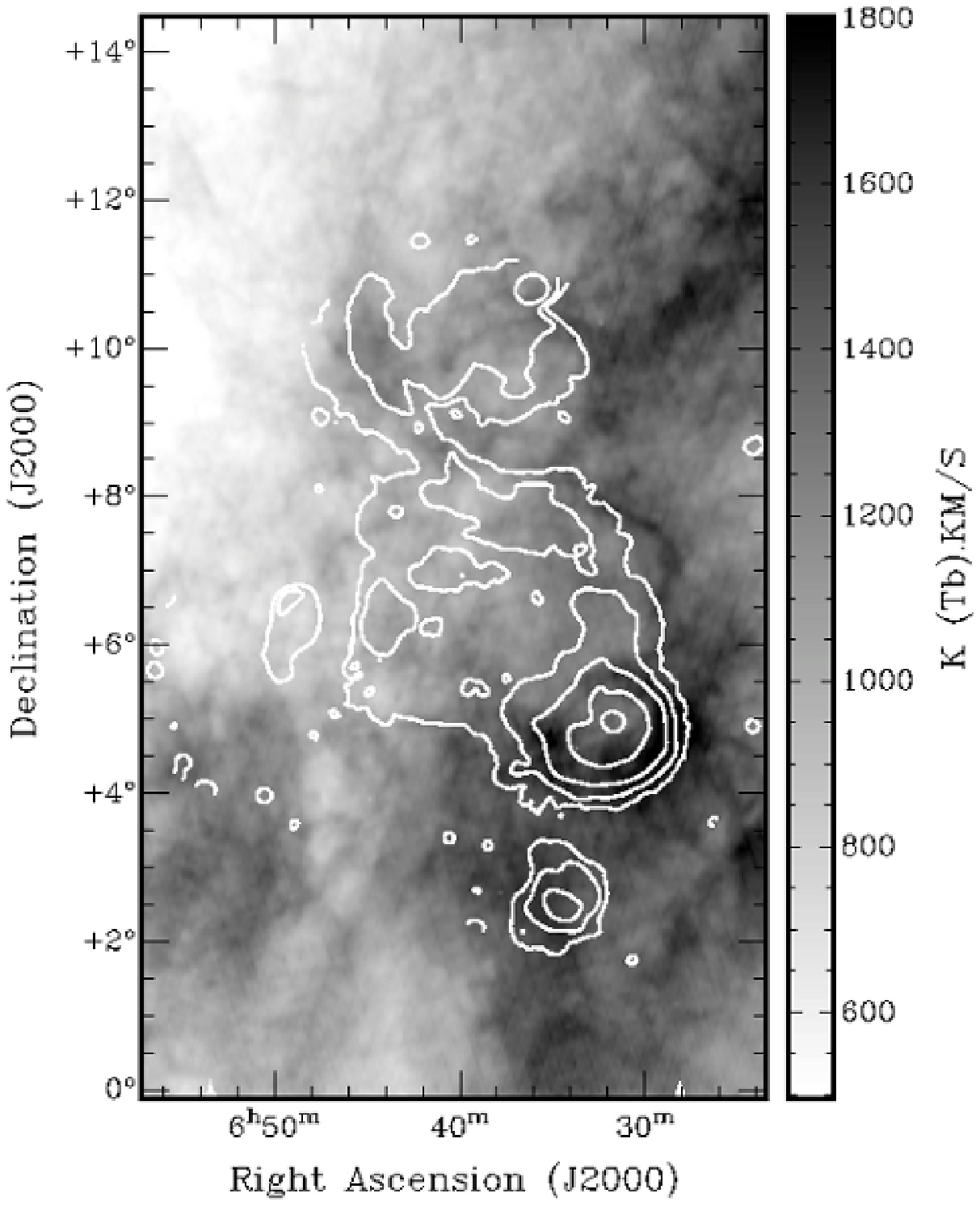}
\caption{Integrated HI-intensities for the velocity range from +4~km~s$^{-1}$ to +25~km~s$^{-1}$ 	with contours of 6~cm continuum emission overlaid, having levels of 10, 30, 120, 		640~mK.
}
\label{HI}
\end{center}
\end{figure}

In Fig.~\ref{HI} we present the result for the integrated HI column densities over the velocity
range $+4$ to $\rm +25~km~s^{-1}$. Weak HI emission segments associated with the eastern radio
shell of the Monoceros SNR emerge after the large-scale diffuse emission has been subtracted
using the ``unsharp-masking'' procedure described by \citet{sr79}. A cavity with an angular diameter of about 3.8$\degr$ is visible coinciding with the Monoceros SNR.

Assuming the HI emission to be optically thin, we calculate the average column density for
the northwest shell to be about $\rm 5.5~\times~10^{20}~cm^{-2}$. If a physical
association exists, with a distance of 1.6~kpc, the thickness of the half-shell of 13$\arcmin$ 
with a 1.9$\degr$ radius corresponds to a swept-up mass ($m_{H}N$(HI)) of about 4140~M$_{\odot}$,
assuming the southern shell has a similar column density to the western shell. The average
density in the HI-shell is then about $n\sim(N/L)=3.6$~cm$^{-3}$ with a line-of-sight depth
within the shell of 49~pc. If this mass of neutral gas was originally uniformly distributed
within a sphere of radius 1.9$\degr$, the pre-explosion ambient density would have been
$\rm 0.27~cm^{-3}$, consistent with the ambient density of $\rm 0.34~cm^{-3}$ estimated
by~\citet{o86}, according to the Chevalier formula~\citep{c74}.
The evacuation by the action of stellar winds within the Mon OB2 association and the progenitor may cause the low density environment. 

\begin{figure}[!hbt]
\begin{center}
\includegraphics[angle=0,width=0.48\textwidth]{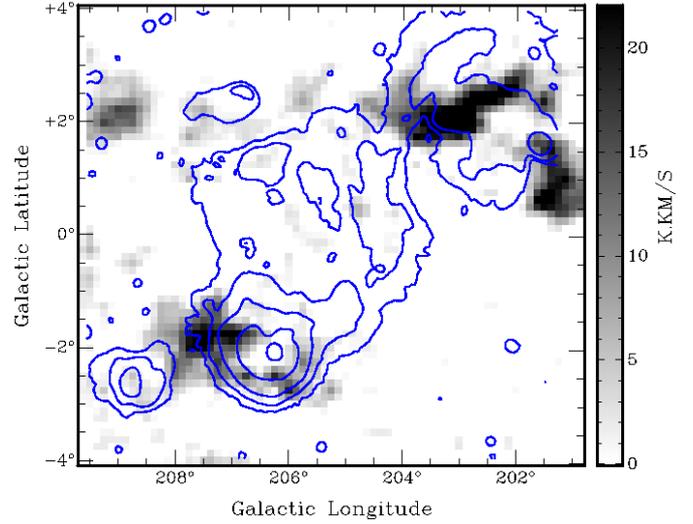}
\caption{Integrated CO-intensities for the velocity range from 0~km~s$^{-1}$ to +30~km~s$^{-1}$ with 6~cm continuum contours overlaid, having levels of 10, 30, 120, 640 mK.
}
\label{co}
\end{center}
\end{figure}

\subsection{CO associated with the Monoceros SNR}
We inspected the CO images of this region obtained by~\citet{omt96}. There are weak CO clouds in
the direction of the enhanced radio continuum emission of the western shell with velocities
ranging from $-$3.1~km~s$^{-1}$ to +9.4~km~s$^{-1}$, and enhanced CO emission is found towards
the southern shell with velocities ranging from~$+10$ to $+30$~km~s$^{-1}$.
The Cone cloud adjacent to the northwest of the Monoceros SNR has an average velocity
of +7~km~s$^{-1}$. This cloud is the progenitor of the young open cluster NGC~2264
and belongs to the Mon OB1 association~\citep{omt96}.

In Fig.~\ref{co} we show the integrated $^{12}$CO emission from~\citet{omt96} in the velocity
interval $0<v<+30$~km~s$^{-1}$ with contours of the 6~cm continuum emission overlaid.
The molecular cloud associated with the Rosette Complex has a ring shape around the nebula with a
gap towards the Monoceros SNR. Based on the associated CO morphology and the fact that the Monoceros SNR has a similar velocity to that of the Rosette Molecular Complex, we suggest
that the Monoceros SNR is interacting with the Rosette Complex, and that the progenitor of
the Monoceros SNR comes from the Rosette Complex.

\subsection{Interaction with the Rosette Nebula}
The high resolution HI channel maps provide evidence that the Monoceros SNR is interacting
with the Rosette Nebula. In the integrated HI emission map of Fig.~\ref{HI}, the Rosette Nebula
appears to have broken into the Monoceros SNR area outlined by its outer HI shell. 
the HI shell associated with the Rosette Nebula presents a ring shape with a minimum
in the direction of the Monoceros SNR. In the channel maps of Fig.~\ref{channel}, a small section
of possible HI shell structure is seen at $v\sim+12$~km~s$^{-1}$ projecting from the east
of the Rosette Nebula, coinciding closely with the possible new radio shell.
This feature might have a relationship with the interaction between the Monoceros SNR and the
Rosette Nebula, and is possibly caused by the shock of the SNR or stellar winds in the Rosette
Nebula.
 
\begin{figure*}[!hbt]
\begin{center}
\includegraphics[angle=-90,width=0.48\textwidth]{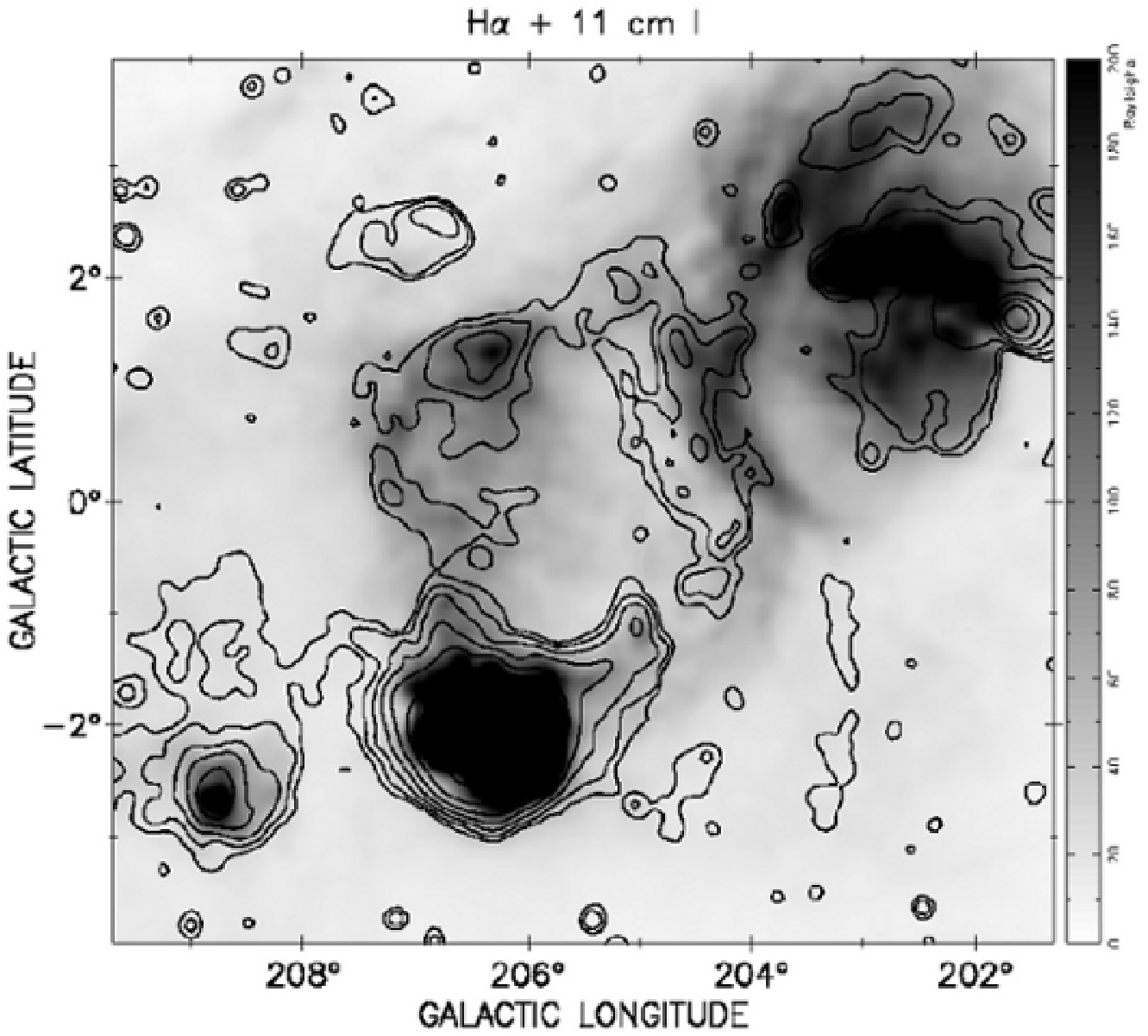}
\includegraphics[angle=-90,width=0.48\textwidth]{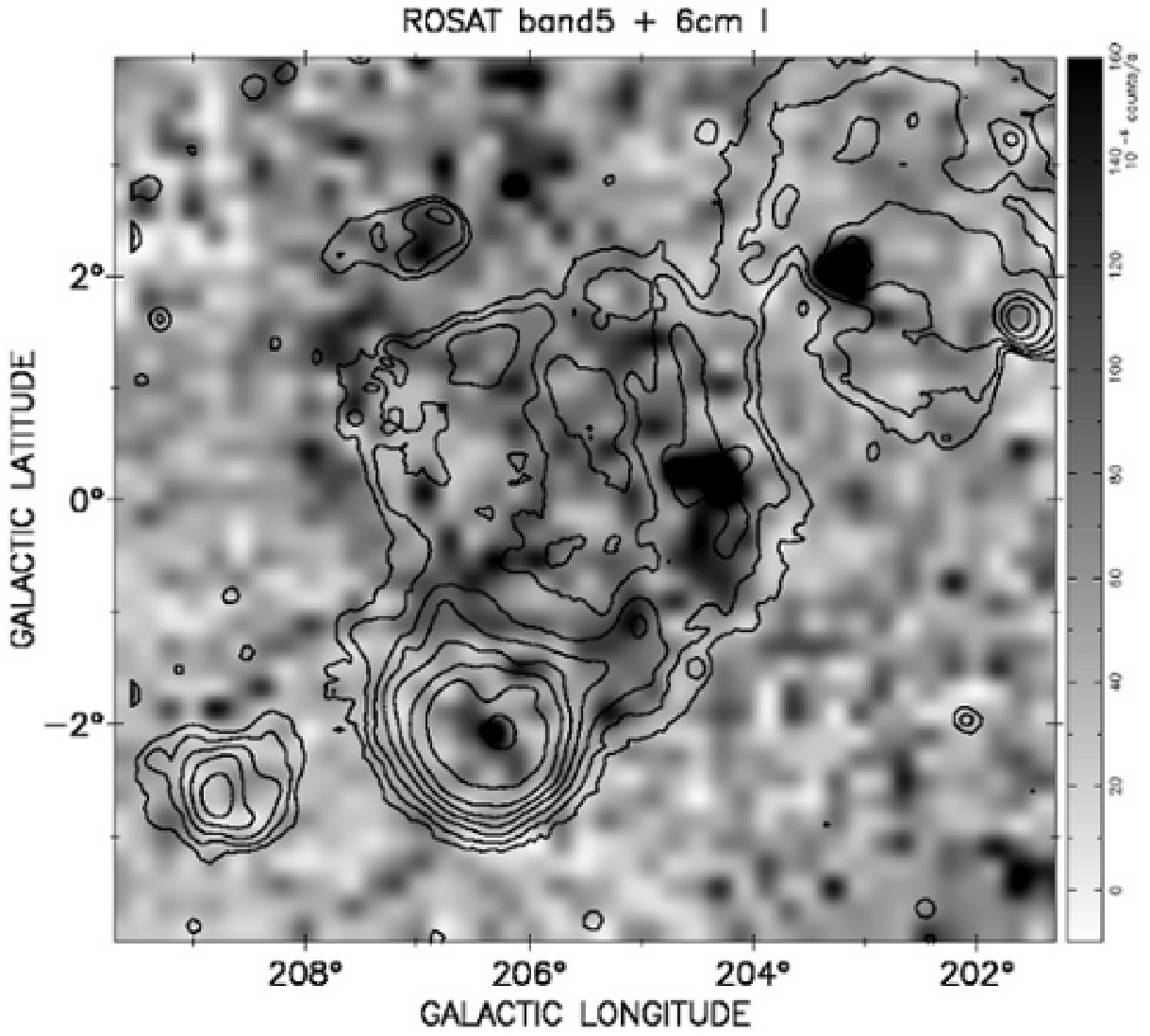}
\caption{Upper panel: H$\alpha$ emission overlaid with 11~cm continuum
     contours having the same levels as for Fig.~\ref{cont}. Lower panel: ROSAT X-ray
     emission smoothed to 10$\arcmin$ angular resolution overlaid with 6~cm total
     intensity contours haveing the same levels as for Fig.~\ref{6cm}.
}
\label{multi}
\end{center}
\end{figure*}

It has been previously speculated that the formation of the cluster NGC~2244 was triggered by an
encounter between the Monoceros SNR and the RMC~\citep{tfm03}. Morphologically, it seems that the
progenitor of the SNR exploded first and its shock formed a well-shaped shell.
The expansion of the Rosette Nebula ionized by the cluster NGC~2244 collided with the
SNR shell and kept expanding. However, the SNR's age has been estimated to be
0.3$-$5$\times 10^{5}$~yr~\citep{wss01}. The main-sequence turnoff age of the cluster is about
1.9~Myr~\citep{ps02}, which suggests that the cluster formed earlier than any possible encounter
with the SNR. The kinetic energy of the expansion of the Rosette Nebula was about 
$\sim3.8\times10^{48}$~ergs~\citep{kb93}, which is smaller than the typical SNR explosion energy
of $10^{51}$~ergs. Although there is little nonthermal radio emission detected in the Rosette
Nebula, it is still possible that the shock of the Monoceros SNR has broken into the Rosette
Nebula at some places, possibly the new radio shell region, and is triggering new star formation
in the cluster.  

Despite the fact that no detection of maser emission has been reported in the radio survey at
1720.5~MHz by~\citet{fgr96}, the Monoceros SNR has been interpreted as interacting with the
atomic and molecular components of the Rosette Complex environment
based on enhanced line widths of H109$\alpha$ and H$\alpha$ emission~\citep{fgo79},
and similar velocity of H$\alpha$ emission in the contact region~\citep{r83}.
EGRET and HESS have detected high energy $\gamma$-ray emission from the overlap region
between the Monoceros SNR and the Rosette Nebula, interpreted in terms of $\pi^{0}$ decay
resulting from collisions between SNR-accelerated particles and the molecular cloud
~\citep{jbd97,trd03}. For the first time, the present HI observations provide direct morphological
evidence of their interaction. The Monoceros SNR has swept up an HI mass of
about 4000~M$_{\odot}$, with a possible break in the shell at +12~km~s$^{-1}$ towards the
Rosette Nebula. A number of Galactic SNRs in physical contact with molecular clouds have been
summarized by~\citet{jcw10}. A similar case is the SNR HB3, which is interacting with
the adjacent HII region W3 and the associated star-formation region~\citep{rdl91}.

\section{Comparison with other multi-bands images}
We compare multi-wavelength observations to discuss the physical properties and the
environment of the Monoceros SNR.

The Monoceros SNR shows diffuse H$\alpha$ emission associated with most of the SNR shell
region, and outside of the shell to the southwest shell (Fig.~\ref{multi}). 
The diffuse emission in the area east of the SNR represents a low density environment, 
which results in a less compressed magnetic field and thus reduced efficiency of particle
acceleration. The radio filaments are well correlated with the optical filaments
([NII])~\citep{deg78}, except for the inner western filament. This correlation is not seen in the
optical image, probably due to lower temperatures/densities in the radio filaments. 

The Monoceros IRIS 60~$\mu$m dust emission with 6~cm total intensity contours superposed is shown
in Fig.~\ref{dust}. The shell-associated dust emission stretching from the eastern spur of the
Rosette Nebula is enhanced with high ISM density. The dust emission envelope in the western region
is outside the radio shell, and is associated with the diffuse optical shell~\citep{deg78}.
It is also well circumscribed by the HI shell, suggesting that the dust there has been
ionized by a pre-shock caused by stellar winds from the progenitors.

The Monoceros image (0.56-1.21~keV) from the ROSAT soft X-ray all-sky survey~\citep{sef97}
is shown in Fig.~\ref{multi} with contours of 6~cm total intensity overlaid.
Diffuse X-ray emissions is located on the rim of the remnant and coincides with
the densest region of optical filaments, as expected for a young SNR in the Sedov phase.
In the eastern SNR shell, the X-ray emission is located outside the boundary of the radio contours  which are assumed to trace the outer SNR shock, indicating more diffuse emission in the weak eastern shell.

\citet{lns85,lns86} fitted the weak X-ray emission detected by the Einstein X-ray satellite, and
concluded that the initial density of the medium in which the Monoceros supernova occurred was
0.003~cm$^{-3}$, and the expansion of the SNR proceeds in a non-homogeneous multi-component medium.
The HI environment confirms this and shows a gradient of density.
The Monoceros SNR has probably expanded in a pre-existing cavity generated by the progenitor.
In this case, the cavity would have had to be extremely large, about 105~pc, which may
have required more than one star to make. The shock has recently encountered the
interstellar material, which turns it into a bright emitter at radio wavelengths.

\section{Summary}
After subtraction of the inner thermal emission and compact extragalactic sources,
we obtained integrated flux densities of 120.8$\pm$9.9~Jy at 21~cm, 89.7$\pm$7.9~Jy at
11~cm and 73.4$\pm$3.5~Jy at 6~cm for the Monoceros SNR. 
The spectral index from these three wavelengths is about $\alpha=-0.41\pm0.16$,
which was confirmed by TT-plots. A new southern shell branch is found, likely
belonging to the Monoceros SNR, based on its spectral index of $\sim -0.5$.
The distribution of spectral index over the SNR shows some variations,
and steepens towards the western inner filamentary region, corresponding to a weaker
shock compression.

Strong polarized emission is observed from the western filament of the Monoceros SNR
at 6~cm and 21~cm, indicating the presence of a strong regular magnetic field component.
The RM values there are ambiguous, and are typically 30$\pm$77n~rad~m$^{-2}$,
with $n=1$ being the favored value. The magnetic field in the western shell region is estimated
to be 9.5~$\mu$G.

The case for interaction between the Monoceros SNR and the Rosette Nebula is strengthened by
the HI data. We identify partial neutral hydrogen shell structures at LSR velocities of
+15~km~s$^{-1}$ circumscribing the continuum emission. The HI shell has swept up a mass of
about 4000~M$_{\odot}$ for a distance of 1.6~kpc. The Monoceros SNR has probably triggered
part of the star formation in the Rosette Nebula. The western HI shell is found to be well
correlated with the dust emission outside of the radio shell, indicating that the Monoceros SNR
is evolving within a cavity blown-out by the progenitor, and that the SNR is still in the
Sedov phase.

\begin{acknowledgements}
The 6~cm data were obtained with a receiver system from the MPIfR mounted at the
Nanshan 25~m telescope at the Urumqi Observatory of NAOC. We thank 
the helpful Urumqi staff and people in the Sino-German 6~cm survey team for their
contribution to the hard work. The GALFA-HI survey is obtained with a 7-beam feed array
at the Arecibo Observatory. We are grateful to the staff at the Arecibo Observatory,
as well as the GALFA-HI survey team for conducting the GALFA-HI observations.
This research was supported by the National Science Foundation of China (grant 11073028). M. Z. acknowledges support from the ``Hundred-talent program'' of the Chinese Academy of Sciences. We thank Dr. Youling Yue and Dr. Lei Qian for useful discussions
and helpful instruction during this work. We thank the anonymous referee for his instructive and useful comments. We also thank Dr. James Wicker for proof reading the manuscript.

\end{acknowledgements}

\bibliographystyle{aa}

\bibliography{Mon-bb}

\end{document}